\documentclass[12pt,preprint]{aastex}

\usepackage{emulateapj5}

\newcommand{\et}{et al.}

\newcommand{\fv}{F_{var}}
\newcommand{\fvar}{F_{var}}

\newcommand{\mbh}{M_{{\rm BH}}}
\newcommand{\fracthefsh}{\frac{F_{var,soft}}{F_{var,hard}}}
\newcommand{\Dtsamp}{{\Delta}T_{{\rm samp}}}

\newcommand{\xte}{{\it RXTE}}

\newcommand{\asca}{{\it ASCA}}
\newcommand{\Msun}{\hbox{$\rm\thinspace M_{\odot}$}}
\def\lineindex#1{{\thinspace\small#1}}
\def\I{\lineindex I}  
\slugcomment{}
\lefthead{Markowitz \& Edelson}
\righthead{Seyfert 1 X-ray Variability}

\begin{document}
\title{An Expanded {\it RXTE} Survey of X-ray Variability in Seyfert 1 Galaxies}

\author{A.~Markowitz\altaffilmark{1,2}, R.~Edelson\altaffilmark{3}
\altaffiltext{1}{Laboratory for High-Energy Astrophysics, NASA/Goddard Space Flight Center, Code 662, Greenbelt, MD 20771; agm@milkyway.gsfc.nasa.gov}
\altaffiltext{2}{N.A.S./N.R.C.\ Research Associate}
\altaffiltext{3}{Dept.\ of Astronomy, Univ.\ of California, Los Angeles CA
90095-1562}}

\begin{abstract}

The first seven years of {\it RXTE} monitoring of Seyfert~1 
active galactic nuclei have been systematically analyzed to yield
five homogeneous samples of 2--12~keV light curves, probing 
hard X-ray variability on successively longer durations from 
$\sim$1 day to $\sim$3.5 years. 2--10~keV variability on time scales 
of $\sim$1 day, as probed by {\it ASCA}, are included. All sources 
exhibit stronger X-ray variability towards longer time scales,
but the increase is greater for relatively higher luminosity sources.
Variability amplitudes are anti-correlated with X-ray luminosity
and black hole mass, but amplitudes 
saturate and become independent of luminosity
or black hole mass towards the longest time scales. The data are 
consistent with the models of power spectral density (PSD) movement 
described in Markowitz \et\ (2003a) and McHardy \et\ (2004), whereby 
Seyfert 1 galaxies' variability can be described by a single, 
universal PSD shape whose break frequency scales with black hole mass.
The best-fitting scaling relations between variability time scale, 
black hole mass and X-ray luminosity imply an average accretion rate 
of $\sim$5$\%$ of the Eddington limit for the sample.
Nearly all sources exhibit stronger variability in the relatively soft 
2--4 keV band compared to the 7--12 keV band on all time scales. 
There are indications that relatively less luminous or less massive
sources exhibit a greater degree of spectral variability for a given 
increase in overall flux. 

\end{abstract}


\section{Introduction}

X-ray observations can provide constraints on the physical conditions
in the innermost regions of Seyfert~1 Active Galactic Nuclei (AGNs),
as the X-rays are generally thought to originate in close proximity 
to the putative central supermassive black hole. On the basis of 
spectroscopic observations, the leading models of the X-ray continuum 
production are based on a hot, Comptonizing electron or electron-positron 
pair corona close to the black hole. The exact geometry remains uncertain, 
though numerous models have been invoked (e.g, Zdziarski \et\ 2003), 
including a neutral accretion disk extending in to the minimum stable 
orbit and sandwiched by a patchy and possible outflowing corona 
(e.g., Stern \et\ 1995, Svensson 1996, Beloborodov 1999) and a hot inner 
disk radially surrounded by a cold disk, with a variable transition 
radius (Shapiro, Lightman \& Eardley 1976, Zdziarski, Lubinski \& Smith 1999). 
The corona multiply-upscatters thermal soft photons emitted from the disk 
to produce an X-ray power-law in the energy range 1--100~keV (e.g., 
Haardt, Maraschi \& Ghisellini 1994). Furthermore, the disk, or some other 
cold, optically thick material, reprocesses the hard X-rays, as 
evidenced by the so-called 'Compton reflection humps' above $\sim$10~keV in 
Seyfert spectra, as well as strong iron fluorescent lines at $\sim$6.4~keV
(Lightman \& White 1988, Guilbert \& Rees 1988, Pounds \et\ 1990).

Seyfert~1 galaxies exhibit rapid, aperiodic X-ray continuum variability for 
which no fully satisfying explanation has been advanced. Probably the best 
way to characterize single-band Seyfert variability, if adequate data
exist, is to measure the fluctuation power spectral density (PSD) function.
Recent studies such as Edelson \& Nandra (1999), Uttley, McHardy \& 
Papadakis (2002), Markowitz \et\ (2003a), Marshall \et\ (2004) and McHardy 
\et\ (2004) measured high-dynamic range broadband PSDs which showed the 
red-noise nature of Seyfert variability at high frequencies, but flattened 
below temporal frequencies corresponding to time scales of a few days. 
Markowitz \et\ (2003a) developed a scenario in which all Seyfert~1s have a 
PSD shape similar to that of X-ray Binaries (XRBs) and which scale 
towards lower temporal frequency with increasing black hole mass. 
Physically, this is consistent with a scenario in which relatively 
more massive black holes host larger X-ray emitting regions, the 
variability mechanism takes a longer time to propagate through the 
emission region, and the observed variability is 'slower.' 

When data are not adequate to construct a PSD, it is still valuable to
quantify the variability amplitude. The well-known anticorrelation between 
variability amplitude (as quantified over a fixed temporal frequency range) 
and source luminosity on both short time scales ($\sim$1~d: Barr \& 
Mushotzky 1986; Nandra \et\ 1997a, Turner \et\ 1999) as well as long time 
scales ($\sim$300~d: Markowitz \& Edelson 2001, hereafter ME01) is 
consistent with the above physical interpretation.

Numerous X-ray spectral variability studies (e.g., Markowitz, Edelson 
\& Vaughan 2003b; also Nandra \et\ 1997a, ME01) have shown the majority 
of Seyferts to soften as they brighten, with the relatively softer 
energies displaying stronger variability. It is currently unclear 
whether this is due to intrinsic slope changes of the coronal power-law 
continuum (e.g., Lamer \et\ 2003a, Uttley \et\ 2003) or due to the 
presence of a much less variable hard component that is likely associated 
with the Compton reflection hump (e.g. Shih, Iwasawa \& Fabian 2002; 
Taylor, Uttley \& McHardy 2003). In contrast to the 'normal' or 
'broad-line' Seyfert 1's which show this property, however, some 
'Narrow-Line' or 'soft-spectrum' Seyfert 1s (characterized by FWHM $<$ 
2000 km s$^{-1}$, and steep photon indices; e.g., Boller, Brandt \& 
Fink 1996) have been seen to vary with a much weaker dependence on 
energy compared to broad-line Seyfert 1s (e.g., Edelson \et\ 2002, 
Vaughan \et\ 2002).

The archival data accumulated by the {\it Rossi X-ray Timing Explorer} 
({\it RXTE}) during its first seven years of operation permits a study of 
broadband continuum and spectral variability behavior on time scales 
ranging from days to years. The long-term variability survey of ME01 was 
the first to systematically probe X-ray variability on such long time 
scales, examining nine Seyfert~1 light curves each of 300 days in duration. 
This paper expands that survey to cover additional time scales and sources
using additional archival {\it RXTE} data. In this paper we test the 
relation between X-ray variability and black hole mass, including the idea 
of broadband PSD movement with black hole mass, and exploring spectral 
variability throughout Seyfert~1s. The source selection and data reduction 
are described in $\S$2. The sampling and analysis are described in $\S$3.
The results are discussed in $\S$4, and a short summary is given in $\S$5.
An Appendix briefly explores if the modeled {\it RXTE} PCA background has 
any significant effect on the measured variability properties for low count 
rate or steep-spectrum sources.

\section{Data Collection and Reduction}

{\it RXTE} has observed $\sim$55 Seyfert~1 galaxies during the first seven 
years of its mission. Data taken through most of Cycle 7 had turned public
by 2004 February, when these analyses were performed. This paper considered 
these data as well as the authors' proprietary observations of three Seyfert~1
galaxies observed during Cycle 8. $\S$2.1 details how the {\it RXTE} data 
were reduced.

The observational approach of this project was to obtain monitoring on 
multiple long time scales, sampled as uniformly as possible for as many 
Seyfert~1 galaxies as possible. Using the available archive of \xte\ data 
to optimize this trade-off yielded a sample of 27 Seyfert~1s suitable for 
analysis on at least one of the time scales of interest, 1 d, 6 d, 36 d, 
216 d, or 1296 d. Additionally, most of these sources also had adequate 
short time scale (1 d) {\it ASCA} data publicly available. Most of the 
sources with data on the 36 d, 216 d, and 1296 d time scales have had 
their PSDs measured or are currently undergoing monitoring for future PSD 
measurement. $\S$2.2 and $\S$2.3 detail construction of the {\it RXTE} 
and {\it ASCA} light curves, respectively.

\subsection{{\it RXTE} data reduction}      

All of the {\it RXTE} data were taken with the Proportional Counter Array 
(PCA), which consists of five identical collimated proportional counter 
units (PCUs; Swank 1998). For simplicity, data were collected only from 
those PCUs which did not suffer from repeated breakdown during on-source time
(PCUs 0, 1, and 2 prior to 1998 December 23; PCUs 0 and 2 from 
1998~December~23 until~2000~May~12; PCU~2 only after 2000 May 12). Count 
rates quoted in this paper are normalized to 1 PCU. Only PCA STANDARD-2 data 
were considered. The data were reduced using standard extraction methods and 
{\sc FTOOLS~v5.2} software. Data were rejected if they were gathered less than 
10$\arcdeg$ from the Earth's limb, if they were obtained within 30~min after 
the satellite's passage through the South Atlantic Anomaly (SAA), if 
{\sc ELECTRON0}~$>$~0.1 ({\sc ELECTRON2} after 2000~May~12), or if the 
satellite's pointing offset was greater than 0$\fdg$02.
 
As the PCA has no simultaneous background monitoring capability, background 
data were estimated by using {\sc pcabackest~v2.1e} to generate model files 
based on the particle-induced background, SAA activity, and the diffuse X-ray 
background. This background subtraction is the dominant source of systematic 
error in \xte\ AGN monitoring data (e.g., Edelson \& Nandra 1999). 
Unmodelled variations in the instrument background are usually 
less than 2 percent of the total observed (sky plus instrument) 
background at energies less than 10 keV (Jahoda et al., 
in prep.\footnote{see also
http://lheawww.gsfc.nasa.gov/users/craigm/pca-bkg/bkg-users.html}).
Ignoring the statistical uncertainty (there was adequate 
signal-to-noise in all observations), a systematic uncertainty 
of $\lesssim$2 per cent should thus be kept in mind for all fluxes.
Counts were 
extracted only from the topmost PCU layer to maximize the signal-to-noise 
ratio. All of the targets were faint ($<$~40~ct~s$^{-1}$~PCU$^{-1}$), so the 
applicable 'L7-240' background models were used. Because the PCU gain settings 
changed three times since launch, the count rates were rescaled to a common
gain epoch (gain epoch 3) by calibrating with several public archive Cas~A 
and Crab observations. Light curves binned to 16~s were generated for all 
targets over the 2--12~keV bandpass, where the PCA is most sensitive and the
systematic errors and background are best quantified. Light curves were also 
generated for the 2--4 and 7--12 keV sub-bands. The data were binned on the 
orbital time scale; orbits with less than ten 16-second bins were rejected. 
Errors on each point were obtained from the standard deviations of the data 
in each orbital bin. Further details of \xte\ data reduction can be found in 
e.g., Edelson \& Nandra (1999).

\subsection{{\it RXTE} sampling}      

The observational approach of this project was to quantify the continuum
variability properties of Seyfert~1 galaxies on multiple time scales. This 
required assembling samples that were, to the greatest degree possible, 
uniformly monitored for proper comparison between sources. Sources with a 
weighted mean count rate significantly below 1~ct~s$^{-1}$~PCU$^{-1}$ over the 
full 2--12~keV bandpass were rejected to minimize the risk of contamination 
from faint sources in the field-of-view, to ensure adequate signal-to-noise,
and to minimize the influence of systematic variations in the modeled X-ray
background.

The sampling of the publicly available data was highly uneven in general. The 
original observations were made with a wide variety of science goals, leading 
to a variety of sampling patterns and durations. This required us to clip 
light curves to common durations and resample at similar rates in order to 
produce samples with homogeneous sampling characteristics. For each total 
light curve, optimum windows of 1 d, 6 d, 36 d, 216 d, and 1296 d 
(evenly-spaced in the logarithm by a factor of 6) were selected. Given the 
original sampling patterns, these windows represented a reasonable spread in 
temporal frequency coverage, and yielded a reasonably-sized sample on each 
time scale. For each time scale, light curves shorter than the optimum window
were rejected. Light curves with long gaps ($>$$1/3$ of the total duration) 
within the window were also rejected. Such gaps reduce the statistical 
significance of parameters derived over the full duration, and interpolating 
across such large gaps would result in an underestimate of the true 
variability amplitude. For each source, as many usable light curves as 
possible on each of the five time scales were selected from the total light 
curves. In NGC~3227, there was a significant hardening of the spectrum during
approximately MJD 51900--52000, consistent with a temporary increase in cold 
absorption due to a dense cloud passing along the line of sight (Lamer, 
Uttley \& McHardy 2003b); these data were excluded.

To extract light curves that were sampled as uniformly as possible, the light 
curves were resampled on each of the five time scales with a common, 
optimized rate. This was done using an algorithm which selected only data 
points in the original light curve that were separated as close as possible to 
the resampling rate $\Dtsamp$, where $\Dtsamp$ was 5760 sec (1 satellite 
orbit), 0.27 d (4 satellite orbits), 1.6 d, 5.3 d and 34.4 d for the 1, 
6, 36, 216, and 1296 d light curves, respectively. Starting with the first 
data point observed (at time $t_1$), the algorithm selected the data point 
observed closest to times $t_2$= $t_1$ + $\Dtsamp$, ... , 
$t_N$ = $t_{N-1}$ + $\Dtsamp$, where N is the number of points in the final, 
resampled light curve. For light curves that were observed with overlapping 
sampling patterns, portions of intense monitoring were not treated differently 
from the rest of the light curve. That is, the algorithm did not do any 
averaging (in bins of size $\Dtsamp$) during times of intensive monitoring, 
as that would yield reduced variability during that period only relative to 
the rest of the light curve. This allowed the entire light curve to sample 
the variability in a uniform fashion. Resampling at rates longer than 
$\Dtsamp$ would have resulted in too few points in each final light curve, 
while resampling at significantly more frequent rates would have resulted in 
light curves that were not sufficiently uniform, given the original range of 
observing patterns. The final light curves were also required to contain at 
least $\sim$20 points  ($\sim$15 on the 1 d time scale) in order to obtain an 
accurate estimate of the variability amplitude as quantified below; those 
light curves with fewer points were discarded. Light curves with poor signal 
to noise (i.e., due to mean count rates significantly less than 1.0) were 
discarded. Given that many sources were observed with overlapping sampling 
patterns, the final light curves for a given source often share data points 
on multiple time scales and are not completely independent.

This reduction yielded a total of 27 sources with sampling on each at
least one of the five \xte\ time scales. This included 
86 observations of 18 sources on the 1 d time scale,
68 observations of 12 sources on the 6 d time scale,
19 observations of 12 sources on the 36 d time scale,
78 observations of 19 sources on the 216 d time scale, and
12 observations of  9 sources on the 1296 d time scale.
Figure 1 shows the full 2--12 keV \xte\ light curves for all 27 sources, 
before resampling, and showing the boundaries of the sampling windows. 
Table 1 lists source observation parameters, including 2--12 keV luminosity 
$L_{2-12}$ and black hole mass estimate $\mbh$, and sampling parameters. 
All source luminosities were calculated using the global mean {\it RXTE} 
count rate and using the HEASARC's online WebPIMMS v.3.4 flux converter 
assuming an intrinsic power-law with a photon index obtained from either 
previously published spectral fits (e.g., Kaspi \et\ 2001, Pounds \et\ 2003) 
or the online TARTARUS database of {\it ASCA} AGN observations (e.g., 
Nandra \et\ 1997a; Turner \et\ 1999). Luminosities were calculated assuming 
$H_{o}$~=~70~km~s$^{-1}$~Mpc$^{-1}$ and $q_{o}$~=~0.5. All black hole mass 
estimates are reverberation-mapped masses from Kaspi \et\ (2000) and 
Wandel, Peterson \& Malkan (1999) except NGC~4051, from Shemmer \et\ (2003), 
NGC~3783, from Onken \& Peterson (2002), NGC~4593, NGC~3516 \& NGC~3227, from 
Onken \et\ (2003), and Mkn~279, from Wandel (2002) and Santos Lleo \et\ 
(2001). Mass estimates for Ark~564, Mkn~766, MCG--6-30-15, MCG--2-58-22 are 
from Bian \& Zhao (2003) and the mass estimate for PKS~0558--504 is from 
Wang \et\ (2001); these latter two works use the empirical Kaspi \et\ (2000) 
relation between optical luminosity and BLR size. No reliable mass estimate 
exists for 3C~111 or IRAS~18325--5926.

\subsection{{\it ASCA} data }

Short-term {\it ASCA} 2--10 keV light curves were obtained from the TARTARUS 
database for the sources with \xte\ data. The count rates in the light curves 
provided had been combined and averaged between {\it ASCA}'s two Solid-state 
Imaging Spectrometers (SIS; Burke \et\ 1994, Gendreau 1995) and binned to 16~s.
For each source, all available light curves longer than 1~d in duration
were selected from the database; otherwise the longest light curve $>$60 ksec
in duration was used. The light curves were binned on orbital time scales, 
yielding 51 light curves of 11--15 consecutive orbital bins for 21 sources.
Background light curves were similarly binned and subtracted to produce net 
count rate light curves. Table~2 lists source observation and sampling 
parameters for the {\it ASCA} data.

\section{Analysis}                      

\subsection{Quantifying variability amplitudes}

Fractional variability amplitudes ($\fv$; e.g., Vaughan \et\ 2003b, Edelson 
\et\ 2002) were measured for each light curve to quantify the intrinsic 
variability amplitude relative to the mean count rate and in excess of the 
measurement noise;
\begin{equation}
\fv = \sqrt{S^2 - \langle \sigma_{err}^2 \rangle \over \langle X
\rangle^2},
\end{equation}
where $S^2$ is the total variance of the light curve, $\langle
\sigma_{err}^2 \rangle $ is the mean error squared and $ \langle X \rangle
$ is the mean count rate of $N$ total points. The error on $\fv$ is
\begin{equation}
\sigma_{\fv} = \sqrt{  \left\{ \sqrt \frac{ \langle\sigma_{err}^2\rangle}{N} \cdot \frac{1}{\langle X \rangle} \right\} ^2         +       \left\{ \sqrt \frac{1}{2N} \cdot \frac {  \langle\sigma_{err}^2\rangle }{\langle X \rangle^2 \fv} \right\} ^2             }
\end{equation}
as discussed in Vaughan \et\ (2003b); this error formulation estimates 
$\sigma_{\fv}$ based on random errors in the data itself, and not due to 
random variations associated with red-noise processes.

In any red-noise stochastic process there will be 
random scatter in independent estimates of the variance 
or $\fv$ over multiple realizations of the process. This 
is a form of "weakly non-stationary" behavior inherent in
red-noise variability processes. Herein, we adopt the 
definition of weak non-stationarity as a description of 
a variability process whose mean and variance show scatter 
over multiple realizations, but whose underlying PSD 
remains constant over time, with expectation values of 
$\fv$, $\langle$$\fv$$\rangle$, remaining constant over 
time as well (e.g., Vaughan 
\et\ 2003b). In other words, while the expectation value 
of the square of $\fv$ is equal to the integrated PSD 
of the underlying variability process, multiple independent 
realizations of that process will yield a range in 
estimates of $\fv$ even if the PSD does not change 
amplitude or shape and $\langle$$\fv$$\rangle$ is 
constant. Factors of 3 or more in the range of $\fv$ 
are not uncommon (e.g., Vaughan \et\ 2003b). Scatter 
in $\fv$ is therefore not necessarily indicative of 
strongly non-stationary behavior. For multiple \xte\ 
or {\it ASCA} light curves for a given source and time 
scale, the values of $\fv$ were averaged, with the 
uncertainty on the average $\fv$ determined statistically. 
However, one needs at least 10--20 independent estimates 
of $\fv$ to adequately test if those estimates are 
consistent with their expectation value $\langle$$\fv$$\rangle$ 
(see Vaughan \et\ 2003b for detailed descriptions of such 
tests). There are only three objects with enough data for 
this relatively strong test, NGC~7469, IRAS~18325--5926
and MCG--6-30-15 on the 1~d time scales (with the \xte\ 
2--12 keV and \asca\ 2--10 keV values considered together); 
in all three cases at least 70$\%$ of the individual values 
of $\fv$ are consistent with $\langle$$\fv$$\rangle$. 
For the rest of the sample, when multiple estimates of $\fv$ 
were made, the measured values were usually reasonably 
close to $\langle$$\fv$$\rangle$. This is consistent 
with weakly non-stationary behavior. Thus, these values 
of $\fv$ are used hereafter. 

A linear relation between absolute rms variability 
amplitude and flux has been observed in XRBs (Uttley \& 
McHardy 2001) and Seyfert~1s (Edelson \et\ 2002, Vaughan, 
Fabian \& Nandra 2003a). This is a form of non-stationary 
behavior (independent of the weak non-stationarity 
discussed above), as the expectation value of the variance 
is not constant over time. Quantifying variability using 
$\fv$ removes this trend; since the rms--flux relation 
slope is generally seen to be close to 1, $\fv$ thus would
be independent of flux level in the absence of additional 
sources of non-stationarity.

Table~3 lists $\fv$ for each \xte\ light curve over
the 2--12 keV, 2--4 keV and 7--12 keV bands. Table~4 
lists $\fv$ measured over the 2--10 keV band for the 
{\it ASCA} data.

\subsection{Construction of correlation diagrams}

Figures~2a and 2b displays the values of the logarithm of $\fv$ plotted 
against $L_{2-12}$ and $\mbh$. $\fv$ itself will not follow a Gaussian 
distribution, but the logarithm of $\fv$ does follow a distribution
that crudely resembles a Gaussian for red-noise PSD slopes of
$=sim$--1 to --2 (see, e.g., Fig.\ 8 of Vaughan \et\ 2003b).
The {\it ASCA} data 
are included and agree well with the 1-d {\it RXTE} data; one should not 
expect any significant difference between parameters derived over the 2--10 
and 2--12 keV bands. Best-fitting power-law slopes for each data set were 
determined using the Akritas \& Bershady (1996) regression,
which accounts for measurement errors and intrinsic scatter; the
slopes are listed in Table~5. The Spearman rank correlation 
coefficients $r$ and probability $P_r$ of obtaining those values of $r$ by 
chance are also listed in Table~5. Also listed in Table 5 is reduced 
chi-squared $\chi^2_r$, calculated using the best-fit power-law. Because the 
uncertainties on $\fv$ were calculated in linear space, when calculating
$\chi^2_r$, the uncertainties in log space were replaced by the average error 
for all five time scales, 0.047 in the log and 0.044 in the log for the 
$\fv$--$L_{2-12}$ and $\fv$--$\mbh$ relations, respectively, to avoid 
unnecessarily weighting $\chi^2$ towards the longer time scales.

For both sets of relations, the values of $r$ are all negative, with the 
absolute values of $r$ decreasing slightly towards longer time scales. Such 
anticorrelations have been observed previously in AGNs for $\sim$1 d time 
scales (Green, McHardy \& Lehto 1993, Nandra \et\ 1997a; O'Neill \et\ 2004). 
However, the slopes and normalizations of the best-fitting logarithmic power 
law for each data set differ: the slopes generally flatten towards longer 
time scales. The 1-d and 216-d time scale $\fv$--$L_{2-12}$ relations are 
generally consistent with the 1 d and 300 d relations of ME01. For both sets 
of relations, for all objects, the values of $\fv$ generally increase towards 
longer time scales, leveling off somewhat beyond approximately the 36 d time 
scale relation, but the highest mass and highest luminosity sources show the 
largest increase. Formally, the fits to all of the best-fit lines are quite 
poor, but using the $\chi^2_r$ values as a measure of intrinsic scatter, the 
1 d \xte\ $\fv$--$\mbh$ relation shows lower scatter than the corresponding 
$\fv$--$L_{2-12}$ relation, but the scatter is greater in the $\fv$--$\mbh$ 
relation in each of the remaining five data sets. The sum of all six 
$\chi^2_r$ values is also greater for the $\fv$--$\mbh$ relations. Within each 
plot, reduced chi-squared tends to decrease towards longer time scales, 
implying greater intrinsic scatter on the shortest time scales. 

It can be seen from the values of $\fv$ listed in Table~3 that most 
observations (56/68) show stronger variability in the 2--4 keV band 
compared to the 7--12 keV band. Formally, the null hypothesis of the 
2--4~keV and 7--12~keV excess variances (square of the $\fv$) being 
consistent is rejected using an F-test at $>$90$\%$ significance in 12 
observations and $>$95$\%$ significance in 8 observations. The correlation 
diagram for 2--4 keV $\fv$ ($F_{var,soft}$) versus 7--12 keV $\fv$ 
($F_{var,hard}$) for all five \xte\ data sets is shown in Figure~3.
$F_{var,soft}$ and $F_{var,hard}$ are well-correlated on all time scales.
However, it can be seen that the vast majority of points lie to the right of 
the dashed line which represents equal variability in the two bands. This 
shows again that most sources exhibit stronger variability in the relatively 
softer band. Values of the Spearman rank correlation coefficient and 
probabilities are listed in Table~5. Best-fitting power-law slopes were 
determined using the Akritas \& Bershady (1996) regression. There is no 
obvious indication that the degree of spectral variability exhibited is 
dependent on the time scale probed. There is no obvious scatter trend with 
time scale, judging from the $\chi^2_r$ values.

Figure~4 shows the ratio of $\fracthefsh$, 2--4 keV $\fv$ / 7--12 keV $\fv$, 
plotted against $L_{2-12}$ for all five \xte\ data sets. The best-fitting 
slopes, obtained using the Akritas \& Bershady (1996) method, and values of 
Spearman rank $r$ and $P_r$ are listed in Table 5. Also listed in Table~5 are 
the slopes and correlation coefficients for $\fracthefsh$ versus $\mbh$ (not 
plotted). The best-fitting slopes are all roughly similar; the slopes might 
be taken as tentative evidence for relatively less luminous or less massive 
sources to be more strongly variable in the soft band. There is no obvious 
indication that the degree of spectral variability exhibited is dependent 
on the time scale probed. Judging from the $\chi^2_r$ values, there is no 
obvious scatter trend with time scale with either $L_{2-12}$ or $\mbh$. The 
respective sums of the five $\chi^2_r$ values are approximately equal, 
implying roughly equal scatter in the $\fracthefsh$--$L_{2-12}$ and 
$\fracthefsh$--$\mbh$ relations.

Color-flux diagrams, in the which the logarithm of the 7--12 keV / 2--4 keV 
count rate hardness ratio (HR) is plotted against the logarithm of the
geometric mean of the count rates in these two bands, are shown in 
Figure 5. To minimize the effects of changes in spectral response due to 
PCA gain epoch changes, only the largest number of points within a single 
gain epoch was selected for each source. Light curves of $\sim$300 days in 
duration, with no resampling, were used; light curves with less than 70 
points were discarded. This yielded a sample of 27 light curves for 14 
sources; date ranges are listed in Table~6. For each source, the data were 
sorted by increasing geometric mean and grouped into bins of 16 points; the 
highest flux bin was ignored if it contained less than 10 points. For most 
sources, the data form a continuous, well-defined region. It is clear from 
these diagrams as well that nearly all sources soften as they brighten. The 
two exceptions, which show either a slight hardening or no spectral 
variability with flux, are the NLSy1 Ark~564 and the radio-quiet quasar
PG~0804+761, as has been reported previously (Edelson \et\ 2002; Papadakis, 
Reig \& Nandra 2003). Also shown in Figure~5 is the best fitting linear fit 
to the binned data. Table 6 lists the mean hardness ratio values 
$\langle$HR$\rangle$ for each source. For the sample as a whole, the 
average of the 27 mean hardness ratios is 1.06. Ten sources' 
$\langle$HR$\rangle$ values are within 20$\%$ of the sample average. 
However, two sources are notably softer, the soft-spectrum source Ark~564 
and the quasar PG~0804+761, both of which are usually measured to have 
relatively steep photon indices (e.g., Leighly 1999; Papadakis, Reig \& 
Nandra 2003). Three sources, NGC~3227, NGC~3516, and NGC~4151, are notably 
harder; these sources are frequently measured to have relatively flat photon 
indices (e.g., Nandra \et\ 1997b; George \et\ 1998).

Also listed in Table 6 is a parameter derived directly from the slope $m$ of 
the best linear fit, $\phi$ = 2.0$^{-m}$, which quantifies the decrease in 
HR for every doubling in geometric mean count rate. Multiply-measured values 
of $\langle$HR$\rangle$ and $\phi$ for a given object tend to be consistent 
with each other, suggesting that sources do not undergo any radical changes 
in spectral variability behavior over times scales of one or two years. 
$\phi$ is greater than 1 for all sources except PG~0804+761 and Ark~564.
It is noted that these two sources have the lowest 7--12 keV mean count rates
in the sample ($\sim$0.3 c/s/PCU). As discussed in more detail in the 
Appendix, it is conceivable that systematic variations in the modeled 
background may contribute greatly to the observed variability at such low 
flux levels, particularly in the 7--12 keV band. The \xte\ data for 
PG~0804+761 and Ark~564 will therefore not be considered further here.

Figure 6 shows $\phi$ plotted against $L_{2-12}$ and $\mbh$. The Spearman 
rank correlation coefficients are given in Table 5. The best-fit lines in 
Figure~6 were again calculated using the method of Akritas \& Bershady (1996).
The best-fitting slopes, listed in Table~5, are negative, indicating that 
relatively less luminous or less massive sources display a stronger degree of 
spectral variability per given increase in overall flux. To estimate the 
scatter, $\chi^2$ was evaluated for each plot and is listed in Table 5; 
formally, the fits to the lines are quite poor, but reduced $\chi^2$ is 
slightly lower in the $\phi$--$\mbh$ plot.

It is noted that other studies (Edelson \et\ 2002; Papadakis, Reig \& Nandra 
2003) have found Ark~564 and PG~0804+761 to show hardness ratios that are 
independent of flux, which would imply values of $\phi$ close to 1. It is 
noted that in both sources, this value would lie reasonably close to the 
observed $\phi$--$L_{2-12}$ anticorrelation. Additionally, the high-mass
PG~0804+761 would lie close to the $\phi$--$\mbh$ anticorrelation; however,
Ark~564 would be a significant outlier if added to the $\phi$--$\mbh$ 
anticorrelation.

\section{Discussion}

When one uses the fractional variability, $\fv$, as a description of the 
intrinsic, underlying variability process, certain caveats must be kept in 
mind when red-noise processes are relevant. Each light curve is an independent 
realization of the underlying stochastic process and there will be random 
fluctuations in the measured variance. However, in the absence of evidence 
for strongly non-stationary behavior in Seyfert light curves (e.g., 
$\S$3.1, Markowitz \et\ 2003a, Vaughan \et\ 2003b), it is assumed hereafter 
that the values of $\fv$ are reasonable quantifications of the intrinsic 
variability amplitude. The reader must keep in mind the previously discussed
limitations when considering such small numbers of $\fv$ estimates.

The anticorrelations between variability amplitude and source luminosity and 
between variability amplitude and black hole mass seen in previous surveys 
are confirmed here on short time scales. In both sets of anticorrelations, 
the best-fitting power-law slopes gradually decrease towards longer time 
scales. The $\fv$ values tend to increase towards longer time scales,
however, they tend to saturate beyond the 36 d time scale. Consequently, the 
increase in $\fv$ is greatest for the higher luminosity sources. As will be 
discussed in $\S$4.1, this trend is consistent with a scaling of PSD break 
frequency with some fundamental parameter, most likely $\mbh$. All of the 
sources exhibit stronger variability towards relatively softer energies. 
Additionally, sub-band $\fv$ values and color--flux diagrams indicate that 
less luminous sources have a tendency to exhibit more spectral variability 
overall. These spectral variability characteristics are discussed in the 
context of simple X-ray reprocessing models in $\S$4.2.

\subsection{The variability--luminosity--$\mbh$ relationship}

Recent PSD studies have yielded PSD breaks on time scales of a few days or 
less; generally, the power-law slopes flatten from $\sim$--2 above the 
break to $\sim$--1 below the break. Markowitz et al.\ (2003a) developed a 
picture in which all Seyfert~1 PSDs have the same shape but whose 
high-frequency break time scales $T_{{\rm b}}$ scale linearly in temporal 
frequency with $\mbh$. This is consistent with observed anticorrelations
between $\fv$ and $\mbh$ (Papadakis 2004, O'Neill \et\ 2004) and $\fv$ and 
X-ray luminosity (e.g., Nikolajuk \et\ 2004). Interestingly, though, the 
PSD break frequencies appeared to be less correlated with 2--10 keV X-ray 
luminosity. There are not adequate data to construct high dynamic range
PSDs for all targets in the current sample. However, it is reasonable to
assume that all Seyferts have similar PSD shapes with breaks. Given the 
ranges of luminosity and black hole masses spanned by the sample, it is 
reasonable to assume that the longest time scales probes in this survey 
are exploring variability on temporal frequencies well below the breaks in 
most or all of the sources. This would then explain why the variability 
amplitudes observed tend to saturate at similar levels on the longest time 
scales probed, strongly reducing the dependence of $\fv$ on $\mbh$ or 
$L_{2-12}$. However, the data are not able to highly constrain if objects' 
PSDs contain a second, low-frequency break, due to the saturation of $\fv$.

Using the values of $\fv$ measured here, it is possible to further test this 
picture. By scaling the break frequency of a broken power-law model PSD with 
$\mbh$, the resulting predicted values of $\fv$ can be compared to the 
observed $\fv$ values on each of the five time scales to quantitatively
constrain the $T_{{\rm b}}$--$\mbh$ relation. Additionally, scaling the 
break frequency of a model PSD with luminosity can similarly constrain the 
relation between $T_{{\rm b}}$ and $L_{2-12}$ (the bolometric and X-ray
luminosities can be related as approximately
$L_{{\rm bol}}$=27$L_{2-12}$; Padovani \& Rafanelli 1988).

For both of these tests, it was assumed that all Seyferts have the same 
singly-broken PSD shape $P(f)$ described by
$P(f) = A(f/f_b)^{-1}$ (for $f < f_b$), or 
$P(f) = A(f/f_b)^{-2}$ (for $f > f_b$). 
$A$ is the PSD normalization at the high-frequency break $f_b$, calculated as 
0.01 (Hz$^{-1}$)/$f_b$, a relation estimated from the $A$--$\mbh$ and 
$T_{{\rm b}}$--$\mbh$ plots of Markowitz \et\ (2003a; their figures 12 and 13).
A linear scaling between $T_{{\rm b}}$ and $\mbh$ (or $L_{2-12}$) was assumed.
The $\fv$ values were calculated by integrating the PSD between the 
temporal frequencies of 1/$D$ (where $D$ is 1, 6, 36, 216, or 1296 days)
and 1/2$\Dtsamp$. The values of $\fv$ measured from the observed light
curves contain additional contributions to the total variability due to 
aliasing, which arises from the non-continuous sampling, and red-noise 
leak, which arises due to the presence of variability on time scales
longer than those sampled. The reader is referred to Uttley, McHardy \& 
Papadakis (2002) or Markowitz \et\ (2003a) for detailed descriptions of 
these distortion effects inherent in PSD measurement and variability 
analysis. The contribution to the total variance from aliasing was estimated 
analytically by integrating the model PSD from a frequency of 1/(2$\Dtsamp$)
to a frequency of 1/(2000~s); no contribution to the aliased power is 
expected from variations on time scales shorter than $\sim$2000 s. Monte 
Carlo simulations were carried out to estimate the contribution to the 
total variance from red-noise leak. For each model PSD, a light curve of 
length 50$D$, where $D$ is the observed light curve duration, was simulated and
split into 50 light curves each of length $D$ to ensure that variability 
power from red-noise leak was present on the same time scales probed by the 
observations (time scales shorter than $D$). The average variance of these 
50 light curves was calculated and compared to the estimated intrinsic 
(i.e., no red-noise leak present) variability estimated above to estimate the 
variability contribution from red-noise leak.

In order to more directly study the link between the $T_{{\rm b}}$--$\mbh$ 
and $T_{{\rm b}}$--$L_{2-12}$ relations and $\fv$, it was necessary to 
remove the influence of the PSD amplitude $A$ at the break on $\fv$.
The accumulation of Seyfert PSDs supports a range in the observed values of 
$A$ (e.g., Uttley et al., in prep.). To remove the dependence of the 
$T_{{\rm b}}$--$\mbh$ and $T_{{\rm b}}$--$L_{2-12}$ relations on $A$, the 
ratios $R_{Fvar}$ of values of 2--12 keV  $\fv$ on six combinations of time 
scales ($R_{1/6}$, denoting $\fv$(1 d) / $\fv$(6 d), $R_{1/36}$, $R_{1/216}$, 
$R_{1/1296}$, $R_{6/216}$, $R_{6/1296}$) were considered. The remaining four 
model ratios are all relatively flat across the ranges of $\mbh$ and
$L_{2-12}$ considered. They do not provide constraints on 
scaling and fitting the model $\fv$ ratio lines, and are
therefore excluded from analysis.

The ratios of predicted $\fv$ values are plotted as solid lines as a 
function of $\mbh$ and $L_{2-12}$ in Figures 7a and 7b, respectively. No 
arbitrary scaling in the y-direction of the resulting values of $R_{Fvar}$ 
was done. Also plotted are the ratios of observed $\fv$ values; observed 
values of $\fv$ on the 1~d time scale were combined between the \asca\ and 
\xte\ data sets by averaging multiple values for each source. The predicted 
$R_{Fvar}$ functions were simultaneously best-fit in the x-direction. The 
fits indicate that the best-fit linear PSD scaling for Figure~7a requires
the relation $T_{{\rm b}}$ (days) = $\mbh$/10$^{6.7}$$\Msun$. The fit is 
formally quite poor, with $\chi^2_r$ equal to 57.1 for 48 degrees of freedom.
For Figure~7b, the linear PSD scaling required is 
$T_{{\rm b}}$ (days) = $L_{2-12}$/ (10$^{43.5}$ erg s$^{-1}$),
with $\chi^2_r$ equal to 490.2 for 48 degrees of freedom.
The modeling is better overall for the $T_{{\rm b}}$--$\mbh$
relation compared to the $T_{{\rm b}}$--$L_{2-12}$ relation,
given the respective values of $\chi^2_r$. These two best-fitting 
relations together suggest that the average accretion rate for the
entire sample is $\sim$5$\%$ of the Eddington limit.

McHardy \et\ (2004) suggested that the normalization of a 
linear $T_{{\rm b}}$--$\mbh$ relation may be dependent 
on some other parameter, possibly the accretion rate.
Under the assumption that the reverberation masses are 
accurate, the picture emerging from PSD measurement 
seems to be revealing a bifurcation in Seyfert PSDs. 
It appears that the PSD breaks of some Seyferts lie 
close to a $T_{{\rm b}}$--$\mbh$ scaling that is 
approximately quantified as $T_{{\rm b}}$ (days) = 
$\mbh$/10$^{6.5}$$\Msun$ (e.g., NGC~3516, NGC~4151,
and NGC~3783; Markowitz \et\ 2003a). This relation
extrapolates 6--7 orders of magnitude to the PSD 
break of Cyg~X-1 in the low/hard state. Other sources 
(NGC~4051 and possibly other Narrow-Line Seyfert~1s, 
McHardy \et\ 2004) seem to require a $T_{{\rm b}}$--$\mbh$ 
scaling that is approximately $T_{{\rm b}}$ (days) = 
$\mbh$/10$^{7.5}$$\Msun$. This relation extrapolates to 
the PSD break of Cyg~X-1 in the high/soft state, arguing 
some connections between these Seyfert~s XRBs in the 
high/soft state. The best-fitting linear $T_{{\rm b}}$--$\mbh$
relation derived from the present sample lies in between 
these two scalings, though much closer to the low/hard state scaling, 
$T_{{\rm b}}$ (days) = $\mbh$/10$^{6.5}$$\Msun$ relation.
This is consistent with the idea that the present 
sample contains a mixture of sources from the two groups,
but the number of sources that scale with Cygnus X-1's 
high/soft state is a small fraction of the whole sample.
Ignoring the five known high/soft state PSD 
sources\footnote{We note that this last analysis 
step refers to high/soft state PSD sources and not
Narrow-Line Seyfert 1s because there may not be a 
one-to-one correspondence between classification 
as a Broad- or Narrow-Line Seyfert and PSD scaling 
category. For example, given their mass estimates,
the PSDs of the Broad-Line Seyfert 1s NGC~3227 and 
Ark~120 are more consistent with scaling with the 
high/soft state of Cygnus X-1 (Uttley et al., in prep.;
Marshall \et\ 2003), while the PSD of the Narrow-Line 
Seyfert 1 Ark~564 is more consistent with scaling 
with the low/hard state of Cygnus X-1. This is why 
NGC~3227 and Ark~120 were included in the high/soft 
PSD scaling category above and Ark~564 was not.}
(NGC~4051; MCG--6-30-15; NGC 3227, Uttley et al., 
in prep.; Ark 120, Marshall \et\ 2003; PG~0804+761, Papadakis,
Reig \& Nandra 2003) indeed gives a slightly lower 
scaling constant in the best-fitting $T_{{\rm b}}$--$\mbh$
relation, $T_{{\rm b}}$ (days) = $\mbh$/10$^{6.6}$$\Msun$
($\chi^2_r$ = 59.4 for 28 degrees of freedom).

As mentioned previously, the large amount of scatter 
inherent in $\fv$ complicates the present analysis. 
Estimates of $\fv$ for a given source will contain scatter 
even when $\langle$$\fv$$\rangle$ is constant, due to 
the stochastic nature of red-noise variability processes. 
Moreover, not all Seyfert PSDs are exactly identical in 
PSD shape and amplitude, meaning that there will be some 
scatter in $\langle$$\fv$$\rangle$ from one object to 
the next, even when $\fv$ is consistently measured over 
identical sampling windows. For instance, fixing the 
high-frequency PSD slope and break frequency while 
doubling the PSD amplitude $A$ will increase $\langle$$\fv$$\rangle$ by 
41 percent for all time scales studied. For fixed $A$ 
and break frequency fixed at 10$^{-6}$ Hz, steepening 
the high-frequency PSD slope from --2.0 to --2.5 will 
decrease $\langle$$\fv$$\rangle$ by 1.4 and 2.2 (decreases of 0.16 and 
0.34 in the log) on the 6 d and 1 d time scales, 
respectively. Finally, the aforementioned bifurcation 
in PSD break frequencies, corresponding to scaling with 
either the high/soft state or low/hard state of Cyg X-1, 
introduces scatter in $\langle$$\fv$$\rangle$ on time scales
longer than $T_{{\rm b}}$. For time 
scales of a year or more, for the black hole masses and 
PSD break frequencies of interest, $\langle$$\fv$$\rangle$ 
will change by a factor of $\sim$3 or more. There hence 
is intrinsic scatter in $\langle$$\fv$$\rangle$ at both 
long and short time scales due to these effects. Assuming 
that the values of $\chi^2_r$ given in Table 5 are an 
adequate characterization of the intrinsic scatter in 
the $\fv$--$L_{2-12}$ and $\fv$--$\mbh$ relations, one 
could speculate that the increased scatter towards shorter 
time scales may indicate that the range of high-frequency 
PSD slopes contributes more to the overall scatter than 
the low-frequency PSD bifurcation. However, removal of 
the five known high/soft state PSD sources fails to 
reduce scatter at long time scales, and it remains 
difficult to identify the dominant source of intrinsic 
scatter.

\subsection{Spectral variability}

The majority of the Seyferts sampled show stronger variability
towards softer energies, as seen from a comparison of the 
2--4 keV and 7--12 keV $\fv$ values, and from the color-flux 
diagrams. Such behavior is consistent with the well-documented 
property of Seyfert~1s to soften as they brighten.
Some works have suggested spectral pivoting of the coronal power 
law about some energy above 10 keV as the explanation for Seyferts' 
softening as they brighten (e.g., Papadakis \et\ 2002). Thermal 
Comptonization models predict changes in the intrinsic spectral slope 
of the coronal component, $\Gamma_{int}$. In the case of a coronal 
cloud that is fed by a variable soft photon seed flux, held at 
constant optical depth, and not pair-dominated, an increase in 
seed flux will lower the electron temperature of the corona and 
steepen the X-ray spectrum (e.g., Maraschi \et\ 1991, Zdziarski 
\& Grandi 2001). Changes in $\Gamma_{int}$ can also arise from 
changes in optical depth (e.g., Haardt, Maraschi \& Ghisellini 
1997), geometry (e.g., Merloni \& Fabian 2001) and energy balance 
(e.g., Zdziarski \et\ 2003). However, spectral variability studies by McHardy, 
Papadakis \& Uttley (1998), Shih \et\ (2002) and Lamer \et\ (2003a) have 
shown that the spectral fit photon index saturates at high flux. 
To explain this effect, McHardy, Papadakis \& Uttley (1998) and 
Shih \et\ (2002) independently proposed the ``two-component'' model 
consisting of a constant hard reflection component superimposed 
upon a soft coronal component that is variable in normalization but 
constant in spectral shape. That is, $\Gamma_{int}$ is constant due 
to both the disk seed and coronal fluxes increasing. As an example, 
a weak dependence of the variability on energy, as has been observed in 
some Narrow-Line Seyfert 1's (e.g., Edelson \et\ 2002), is possible in 
the context of the two-component model if the hard component is absent or 
extremely weak.

The color-flux diagrams not only show that Seyfert~1s generally 
soften as they brighten, they also tentatively suggest that there is more 
spectral variability for a given increase in flux for the relatively less 
luminous, less massive, and more variable overall sources. Additional support 
comes from the marginal anticorrelations between the ratios of the 2--4 keV 
and 7--12 keV $\fv$ and luminosity (Figure~4) and $\mbh$. This trend could be 
due to some variable soft component present in the 2--4 keV band but not 
evident at higher energies; this component could be more prominent or more 
variable in the relatively lower luminosity objects. Alternatively, the 
physical parameters which ultimately constrain the amount of observed 
spectral variability may themselves be more variable in the relatively lower 
luminosity objects.

Another possible contribution to this effect may arise from the energy 
dependence of the high-frequency PSD (e.g., Papadakis \& Nandra 2001, 
Vaughan, Fabian \& Nandra 2003a, McHardy \et\ 2004). At temporal frequencies 
above the break, PSD slopes tend to increase in slope as photon energy 
increases, typically by $\sim$0.1--0.2 for a doubling in photon energy. One 
would then observe a reduction in the ratio of soft to hard X-ray variability 
in more massive or luminous sources, since their PSD breaks appear at 
relatively lower temporal frequencies. However, simulations show that such an 
effect is minor. Simulations of 300-day light curves using PSD shapes with 
energy-dependent high-frequency slopes (change in slope by 0.2 between the 
two bands), energy-dependent normalization $A$ (roughly 50$\%$ higher in the 
soft band; e.g., McHardy \et\ 2004), and a $T_b$--$\mbh$ relation as per above
yield a reduction in the ratio of soft to hard $\fv$ by $\sim$6$\%$ over the 
$\mbh$ range of interest. This corresponds to a change in $\phi$ of only 
$\sim$7$\%$, much smaller than the range observed.

\section{Conclusions}

This paper extends the results of the first long-term X-ray variability
survey of ME01 to additional sources and time scales, including sampling 
variability on time scales well below the putative PSD breaks in Seyferts.
The well-studied luminosity--variability amplitude anticorrelation 
and the anticorrelation between black hole mass 
and variability amplitude are confirmed on short time scales. Variability 
amplitudes increase towards longer time scales, consistent with red-noise 
variability, but the relatively more luminous and more massive
sources show the greatest 
increase. For both sets of anticorrelations, the best-fitting slopes 
decrease towards longer time scales. These trends are consistent with a 
simple scaling of PSD break frequency with black hole mass as suggested by 
Markowitz \et\ (2003a) and McHardy \et\ (2004), with $\fv$ saturating on 
time scales below the PSD breaks. The best-fitting time scale--mass 
relation is quantified as $T_{{\rm b}}$ (days) = $\mbh$/10$^{6.7}$$\Msun$,
and the best-fitting time scale--luminosity relation is quantified as
$T_{{\rm b}}$ (days)  = $L_{2-12}$/(10$^{43.5}$ erg s$^{-1}$), implying an
average accretion rate for the entire sample of $\sim$5$\%$ of the 
Eddington limit. The measurement of a larger number of Seyfert PSDs at low 
temporal frequencies and additional accumulation of $\fv$ measurements 
on multiple time scales for a given object will further clarify the 
relations between PSD break time scale, PSD normalization, $\fv$, black 
hole mass and luminosity.

Nearly all the observations show relatively stronger variability towards 
softer energies, as seen from the values of $\fv$. Color-flux diagrams 
additionally show that sources soften as they brighten. The color-flux 
diagrams also tentatively suggest that sources with relatively lower 
luminosities or black hole masses display a larger range of spectral 
variability for a given increase in total X-ray flux. 

\acknowledgments
The authors thank the referee for a detailed reading of the
manuscript and for providing numerous comments.
The authors acknowledge the dedication of the entire \xte\ mission team.
This work has made use of data obtained through the High Energy
Astrophysics Science Archive Research Center Online Service, provided by
the NASA Goddard Space Flight Center, the TARTARUS database, which is
supported by Jane Turner and Kirpal Nandra under NASA grants NAG~5-7385
and NAG~5-7067, and the NASA$/$IPAC Extragalactic Database which is
operated by the Jet Propulsion Laboratory, California Institute of
Technology, under contract with the National Aeronautics and Space
Administration. 

\appendix
\section{Influence of PCA background modeling on measured variability properties}

Because {\it RXTE} is a non-imaging instrument, the background must be
modeled. However, for very faint targets, including most soft-spectrum 
Seyferts, the estimated PCA background count rate is greater than the 
source count rate. Small systematic errors in the background model will 
thus cause proportionally larger problems for soft-spectrum and low count 
rate sources. In extreme cases, uncertainty in the background model can lead 
to incorrect characterization of the true variability (e.g., the \xte\ 
observation of the soft-spectrum source TON~S180, Edelson \et\ 2002).
This Appendix explores the influence of the background subtraction
by examining the measured variability characteristics as a function of 
source count rate.
                                             
Figure~8 shows the logarithm of 2--12 keV $\fv$ plotted against the
logarithm of the count rate for all sources and time scales (2--10 
keV for the {\it ASCA} data). For multiply-observed sources on each 
time scale, the values of $\fv$ and count rate obtained before 
averaging were used in order to explore the widest range of count 
rates possible. Spearman rank correlation coefficients and 
probabilities are listed in Table~7. Weak to moderate anticorrelations 
are evident on all time scales. These are not the result of any 
correlation between count rate and luminosity. Spearman rank 
correlation coefficients and probabilities are listed in Table~7
for source mean count rate as a function of both $L_{2-12}$ and 
$\mbh$; in general, count rate is seen to be uncorrelated with either
source parameter. The anticorrelations are, however, the result of 
the inclusion of the narrow-line/ soft-spectrum sources. {\it RXTE}, 
lacking coverage below 2~keV, generally cannot observe most 
soft-spectrum sources. Invariably, this class of objects will yield 
lower 2--12 keV count rates compared to normal, broad-line Seyferts. 
However, these objects also tend to be more variable than broad-line 
Seyferts (e.g., Turner \et\ 1999). Recalculation of the Spearman 
rank correlations, excluding the six narrow-line/soft spectrum 
sources, shows the above anticorrelations to be substantially weakened 
on most time scales.  However, on the 6 and 36 d time scales, it 
is necessary to additionally exclude NGC~3227. The 2000 intensive 
monitoring campaign of NGC~3227, from which the 6 d and 36 d light 
curves are derived, happened to catch this highly variable source 
in a relatively low flux state (as shown in Figure~1). Removal of 
these data points further weakens the anticorrelations in those two 
plots. Overall, there is no evidence that the measured variability
characteristics of low count rate sources are 
affected by the \xte\ background modeling on any time scale.

Figure~9 shows the ratio of 2--4 keV $\fv$ / 7--12 keV $\fv$ is plotted 
against the geometric mean of the count rates in these two bands for all 
five \xte\ time scales. Again, for multiply-observed sources on each time 
scale, the values of $F_{var,soft}$, $F_{var,hard}$ and count rate obtained
before averaging were used. The Spearman rank correlation coefficients and 
probabilities are listed in Table~7. The ratio is seen to be generally
independent of count rate for all time scales.

The parameter $\phi$ is plotted against average geometric mean
count rate in Figure 10, and seen to be independent of count rate
for most sources. However, Ark~564 and PG~0804+761, denoted by open circles 
in the figure, have values of $\phi$ less than 1. Systematic background errors 
may be biasing the estimate of the hardness ratio; this may be an artifact 
of the low count rates for these sources, especially in the hard band 
(7--12 keV count rates per PCU are about 0.3 counts sec$^{-1}$ for both 
sources). Spearman rank coefficients and probabilities are given
in Table 7 with and without Ark~564 and PG~0804+761.

\begin{deluxetable}{llccccccc}
\tabletypesize{\footnotesize}
\tablewidth{6.5in}
\tablenum{1}
\tablecaption{Source and \xte\ sampling parameters\label{tab1}}
\tablehead{
\colhead{Time} & \colhead{Source} & \colhead{} & \colhead{log($L_{2-12}$)} & \colhead{log($\mbh$)} & \colhead{MJD} & \colhead{Num.} & \colhead{Mean} & \colhead{Mean}  \\   
\colhead{Scale} & \colhead{Name} & \colhead{z} & \colhead{(erg s$^{-1}$)} & \colhead{($\Msun$)}    & \colhead{Range}   & \colhead{Pts.} & \colhead{c s$^{-1}$} & \colhead{S/N}  }
\startdata
1 d   &PKS~0558--504 & 0.137 & 44.90 & 7.7                    & 50734.31--50735.24 & 14 & 1.29 &  48 \\
      & 3C~111       & 0.049 & 44.63 & ----                   & 50530.93--50531.66 & 11 & 4.36 & 102 \\
      &              &       &       &                        & 51982.37--51983.43 & 13 & 7.95 &  72 \\
      &              &       &       &                        & 51983.43--51984.49 & 12 & 7.94 &  76 \\
      & Mkn~509      & 0.034 & 44.33 & 9.2$\pm$1.1            & 52012.20--52013.26 & 15 & 6.09 &  75 \\
      & 3C~120       & 0.033 & 44.24 & 7.48$^{+0.21}_{-0.28}$ & 52621.91--52622.90 & 12 & 5.47 &  69 \\
      &              &       &       &                        & 52622.90--52623.89 & 13 & 5.61 &  66 \\
      &              &       &       &                        & 52623.89--52624.88 & 13 & 5.98 &  79 \\
\enddata
\tablecomments{The targets are ranked by 2--12 keV luminosity (col.\ [4]).
Redshifts (col.\ [3]) were obtained from the NED database.
References for black 
hole mass estimates (usually the reverberation-mapped 
estimate, e.g., Kaspi \et\ 2000 and Wandel, Peterson 
\& Malkan 1999) are given in $\S$2.2 of the text. 
Col.\ (7) is the number of points in the light curve after clipping and resampling to a common rate.
Col.\ (8) is the mean 2--12 keV count rate per PCU. Col.\ (9) is the signal-to-noise.}
\end{deluxetable}

\begin{deluxetable}{lccccc}
\tablewidth{5.4in}
\tablenum{2}
\tablecaption{\asca\ sampling parameters \label{tab2}}
\tablehead{
\colhead{Source} & \colhead{MJD Date} & \colhead{Sequence} & \colhead{Num.} & \colhead{Mean} & \colhead{Mean}\\
\colhead{Name} & \colhead{Range} & \colhead{ID Number} & \colhead{Pts.} & \colhead{c s$^{-1}$} & \colhead{S/N} }
\startdata
3C~111        & 50126.54--50127.49 & 74087000 & 15 &  0.74  &  34 \\  
PG~0804+761   & 50756.55--50757.55 & 75058000 & 15 &  0.22  &  21 \\  
3C~390.3      & 49307.84--49308.94 & 70005000 & 15 &  0.35  &  24 \\
Mkn~509       & 49471.39--49472.39 & 71013000 & 15 &  1.02  &  39 \\
3C~120        & 49400.66--49401.66 & 71014000 & 15 &  1.06  &  45 \\
MCG--2-58-22  & 49132.89--49133.70 & 70004000 & 13 &  0.24  &  12 \\
              & 50600.47--50601.47 & 75049000 & 15 &  0.71  &  33 \\
              & 50797.98--50798.98 & 75049010 & 15 &  0.70  &  33 \\
\enddata
\tablecomments{Targets are ranked by 2--12 keV luminosity, given in Table~1.
Col.\ (4) is the number of points in the \asca\ light curve after
orbitally binning. Col.\ (5) is the mean count rate averaged between
both SIS instruments. The SIS data were unusable for the observation of 
NGC~4593 and the first observation of NGC~3227; the GIS data were used and 
GIS count rates were converted to SIS count rates using the online W3PIMMS 
tool.}
\end{deluxetable}

\begin{deluxetable}{lccccc}
\tablewidth{6.5in}
\tablenum{3}
\tablecaption{Derived Variability Parameters for the \xte\ light curves \label{tab3}}
\tablehead{
\colhead{Time}  & \colhead{Source} &\colhead{2--12 keV}    &\colhead{2--4 keV  $\fv$ ($\%$)}      & \colhead{7--12 keV $\fv$ ($\%$)} & \colhead{} \\ 
\colhead{Scale} & \colhead{Name}   &\colhead{$\fv$ ($\%$)} &\colhead{($F_{var,soft}$)}              & \colhead{($F_{var,hard}$) } &\colhead{$\frac{F_{var,soft}}{F_{var,hard}}$ }   }
\startdata
1 d & PKS~0558--504 &  12.1 $\pm$ 0.6 & 11.8 $\pm$ 0.9 & 13.9 $\pm$ 1.5 & 0.85 $\pm$ 0.11 \\
    & 3C~111        &   1.1 $\pm$ 0.7 &  0.4 $\pm$ 2.2 &  1.4 $\pm$ 1.4 & 0.31 $\pm$ 1.58 \\
    & Mkn~509       &   1.5 $\pm$ 0.4 &  3.6 $\pm$ 0.8 & Undef.         &                 \\
    & 3C~120        &   4.0 $\pm$ 0.5 &  4.1 $\pm$ 0.9 &  4.5 $\pm$ 0.8 & 0.91 $\pm$ 0.27 \\
    & MCG--2-58-22  &   1.9 $\pm$ 0.4 &  2.9 $\pm$ 0.8 &  1.1 $\pm$ 2.8 & 2.67 $\pm$ 7.01 \\
    & Ark~120       &   6.2 $\pm$ 0.3 &  6.4 $\pm$ 0.5 &  6.8 $\pm$ 0.7 & 0.95 $\pm$ 0.12 \\
    & NGC~5548      &   5.7 $\pm$ 0.2 &  6.0 $\pm$ 0.3 &  5.6 $\pm$ 0.3 & 1.07 $\pm$ 0.08 \\
    & Ark~564       &  18.2 $\pm$ 0.5 & 18.0 $\pm$ 0.7 & 22.6 $\pm$ 1.5 & 0.80 $\pm$ 0.06 \\
\enddata
 \tablecomments{Targets are ranked by 2--12 keV luminosity, given in Table~1.
1~d $\fvar$ for Mkn~279 is undefined at all bands and is not listed.}
\end{deluxetable}

\begin{deluxetable}{lc}
\tablewidth{2.5in}
\tablenum{4}
\tablecaption{Derived Variability Parameters for the \asca\ light curves \label{tab4}}
\tablehead{
\colhead{Source} & \colhead{2--10 keV}\\
\colhead{Name} & \colhead{$\fv$ ($\%$)}}
\startdata
PG~0804+761 & 3.2   $\pm$    1.9  \\
Mkn~509     & 4.0   $\pm$    0.7\\
3C~120      & 1.5   $\pm$    0.9\\
MCG--2-58-22& 3.1   $\pm$    1.0\\
Fairall~9   & 3.4   $\pm$    2.3\\
Ark~120     & 2.7   $\pm$    1.1\\
IC~4329a    & 5.6   $\pm$    0.7\\
NGC~5548    & 6.5   $\pm$    0.9\\
Ark~564     & 17.0   $\pm$    1.2\\
NGC~7469    & 2.2 $\pm$      1.5\\
IRAS~18325--5926& 14.0   $\pm$    1.9\\
NGC~3783    & 6.1   $\pm$    0.8\\
Mkn~766     & 17.6  $\pm$     1.6\\
NGC~4593    & 14.1  $\pm$     0.9\\
NGC~3516    & 10.3  $\pm$     0.9\\
MCG--6-30-15 & 20.0  $\pm$     1.2\\
NGC~4151    &  6.3 $\pm$     0.4\\
NGC~3227    & 14.8  $\pm$     0.9\\
NGC~4051    & 29.6 $\pm$    2.0\\
\enddata
\end{deluxetable}

\begin{deluxetable}{lcccccccc}
\tabletypesize{\footnotesize}
\tablewidth{6.5in}
\tablenum{5}
\tablecaption{Summary of Broadband Variability Correlations \label{tab5}}
\tablehead{
\colhead{Figure}&\colhead{}&\colhead{}&\colhead{Time}&\colhead{No.}& \colhead{}&\colhead{}&\colhead{}&\colhead{}\\
\colhead{Number}&\colhead{x-axis}&\colhead{y-axis}&\colhead{Scale (d)}&\colhead{Pts.}&\colhead{$r$}&\colhead{$P_r$}&\colhead{Slope}&\colhead{$\chi^2_r$} }
\startdata
2a& $L_{2-12}$     &  $\fv$        &1 ({\it ASCA})& 19 & --0.754 & 1.93$\times$10$^{-4}$ & --0.344$\pm$0.011 & 30.7\\
  &                &               &1 ({\it RXTE})& 17 & --0.690 & 2.16$\times$10$^{-3}$ & --0.316$\pm$0.024 & 43.6\\
  &                &               & 6            & 12 & --0.734 & 6.54$\times$10$^{-3}$ & --0.223$\pm$0.012 & 14.3\\
  &                &               & 36           & 12 & --0.531 & 7.54$\times$10$^{-2}$ & --0.176$\pm$0.017 & 15.4\\
  &                &               & 216          & 19 & --0.642 & 3.07$\times$10$^{-3}$ & --0.138$\pm$0.006 & 6.4\\
  &                &               & 1296         &  9 & --0.533 & 0.139                 & --0.135$\pm$0.014 & 8.8\\
2b& $\mbh$         &  $\fv$        &1 ({\it ASCA})& 18 & --0.670 & 2.36$\times$10$^{-3}$ & --0.321$\pm$0.011 & 41.7\\
  &                &               &1 ({\it RXTE})& 15 & --0.839 & 9.14$\times$10$^{-5}$ & --0.368$\pm$0.015 & 18.7\\
\enddata
\tablecomments{ Col.\ (6) is the Spearman rank correlation coefficient.
Col.\ (7) is the probability of obtaining that correlation coefficient 
by chance. The best-fitting slopes, listed in Col.\ (8), were obtained 
using the regression method of Akritas \& Bershady (1996); their bisector 
regression was used for the correlations in Figure 3. The reduced 
chi-squared values $\chi^2_r$ are listed in Col.\ (9). The number of 
degrees of freedom used in calculating $\chi^2_r$ is the number of data 
points (Col.\ [5]) minus 2 for all data sets. For all six data sets in 
Figures~2a and 2b, $\chi^2_r$ was calculated assuming a uniform error bar of 
0.047 and 0.044 in the log, respectively  (one-sided; the average error for 
all data points in the figure). The first value of $\chi^2_r$ listed for 
the Figure 3 correlations was calculated with $F_{var,soft}$ as the 
independent variable and $F_{var,hard}$ as the dependent variable and 
vice-versa for the second value. For the 1 d $\fracthefsh$--$\mbh$ 
correlation, the positive slope is caused by an outlier point, MCG--2-58-22. 
Removal of this points yields $r$=--0.291, $P_r$=0.334, a 
slope of --0.106$\pm$0.024, and $\chi^2_r$=11.0. Ark~564 and 
PG~0804+761 have been excluded from the second row of Figure~6 correlations.}
\end{deluxetable}

\begin{deluxetable}{lccc}
\tablewidth{4.0in}
\tablenum{6}
\tablecaption{Summary of color-flux diagrams \label{tab6}}
\tablehead{
\colhead{Source}&\colhead{MJD}&\colhead{}&\colhead{} \\ 
\colhead{Name}&\colhead{Range}&\colhead{$\langle$ HR $\rangle$}&\colhead{$\phi$}}
\startdata
IC~4329a      & 52737--53035 & 1.10 & 1.178$\pm$0.008\\
NGC~4151     & 51678--51964 & 2.08 & 1.075$\pm$0.003\\
NGC~3783 (2)  & 51976--52272 & 1.16 & 1.207$\pm$0.032\\
NGC~3783 (1)  & 51679--51975 & 1.19 & 1.188$\pm$0.009\\
3C~120        & 52335--52635 & 1.08 & 1.042$\pm$0.005\\
MCG--6-30-15 (3) & 52276--52554 & 0.93& 1.308$\pm$0.010\\
MCG--6-30-15 (1) & 51676--51974 & 0.97& 1.224$\pm$0.015\\
MCG--6-30-15 (2) & 51976--52274 & 0.98& 1.224$\pm$0.006\\
\enddata
\tablecomments{Targets are ranked by mean geometric mean count rate.
Col.\ (3) is the mean 7--12 keV / 2--4 keV count rate
hardness ratio (HR) value for light curve, obtained from the binned data
in Figure~6. Col.\ (4) lists $\phi$, which quantifies the decrease in HR
for every doubling in flux, as described in the text.}
\end{deluxetable}

\begin{deluxetable}{lcccccc}
\tablewidth{6.5in}
\tablenum{7}
\tabletypesize{\footnotesize}
\tablecaption{Summary of Count Rate Correlations \label{tab7}}
\tablehead{
\colhead{Figure}&\colhead{}      &\colhead{}      &\colhead{Time}  &\colhead{No.} &
\colhead{}&\colhead{}  \\
\colhead{Number}&\colhead{x-axis}&\colhead{y-axis}&\colhead{Scale} &\colhead{Pts.}&\colhead{$r$}&\colhead{$P_r$} }
\startdata
8 & Count rate &  $\fv$         & 1 d ({\it ASCA}) (all srcs.) & 48 & --0.240 & 0.100 \\
  &            &                & 1 d ({\it RXTE}) (all srcs.) & 84 & --0.192 & 0.080 \\
  &            &                & 6 d              (all srcs.) & 67 & --0.516 & 7.99$\times$10$^{-6}$ \\
  &            &                & 36 d              (all srcs.) & 19 & --0.577 & 9.67$\times$10$^{-3}$ \\
  &            &                & 216 d             (all srcs.) & 78 & --0.415 & 1.60$\times$10$^{-4}$ \\
  &            &                & 1296 d            (all srcs.) & 12 & --0.816 & 1.21$\times$10$^{-3}$ \\
  &            &                & 1 d ({\it ASCA}) (BL only)   & 31 & --0.202 & 0.276 \\
  &            &                & 1 d ({\it RXTE}) (BL only)   & 56 & --0.308 & 0.021 \\
\enddata
\tablecomments{Col.\ (6) is the Spearman rank correlation coefficient.
Col.\ (7) is the probability of obtaining that correlation coefficient by chance.
For the Figure 10 correlations, Ark 564 and PG~0804+761 have been excluded from the
second row.}
\end{deluxetable}

\clearpage

\begin{figure}
\figurenum{1}
\epsscale{0.94}
\plotone{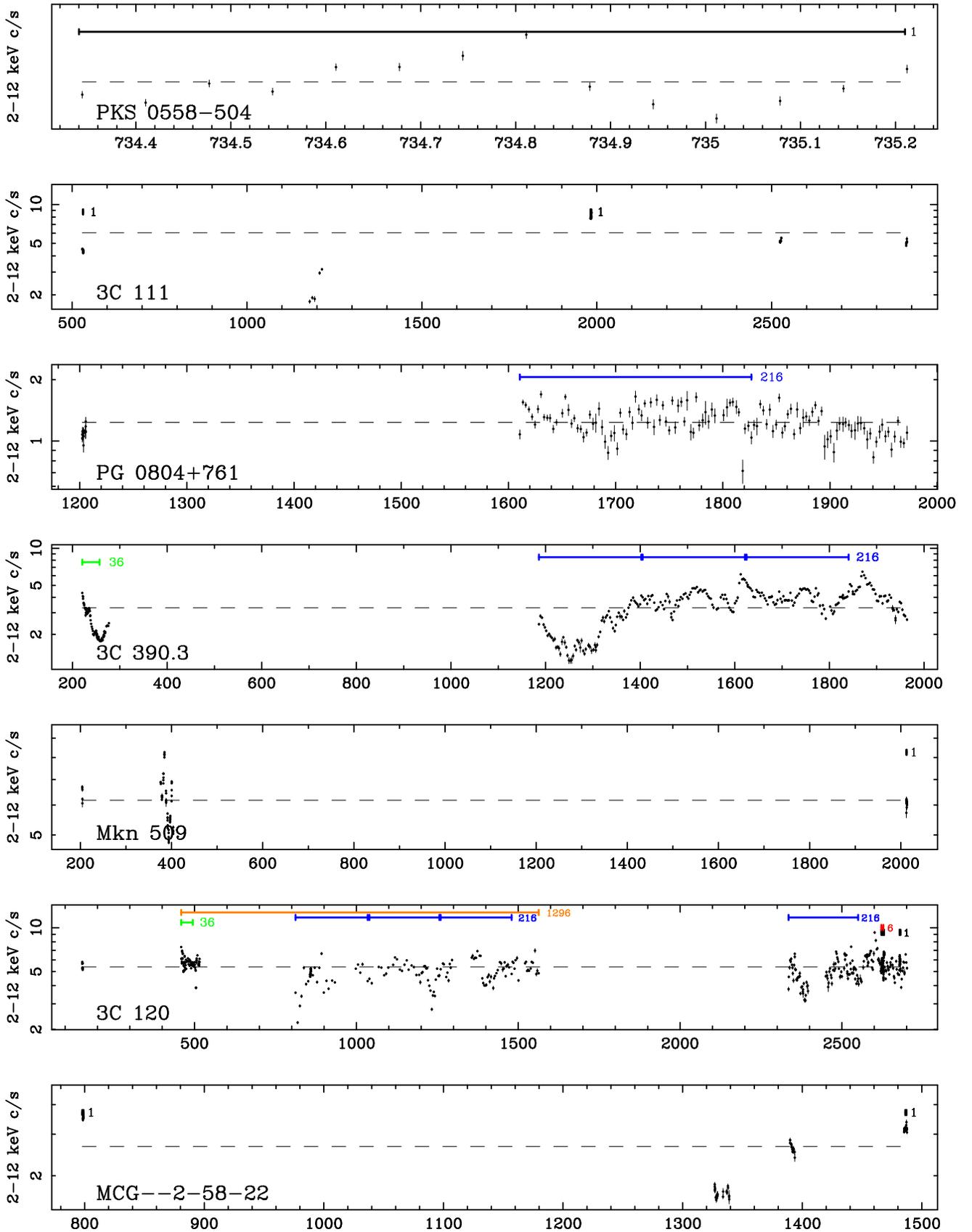}
\caption{ {\it RXTE} 2--12 keV total light curves, ranked by 2--12 
keV luminosity, before clipping and resampling. The black, red, 
green, blue and orange bars denote the extent of the 1 d, 6 d, 
36 d, 216 d, and 1296 d \xte\ light curves, respectively, before
subsampling to a common sampling rate. Error bars are 1$\sigma$.}
\end{figure}

\setcounter{figure}{0}
\begin{figure}
\figurenum{1}
\epsscale{0.94}
\plotone{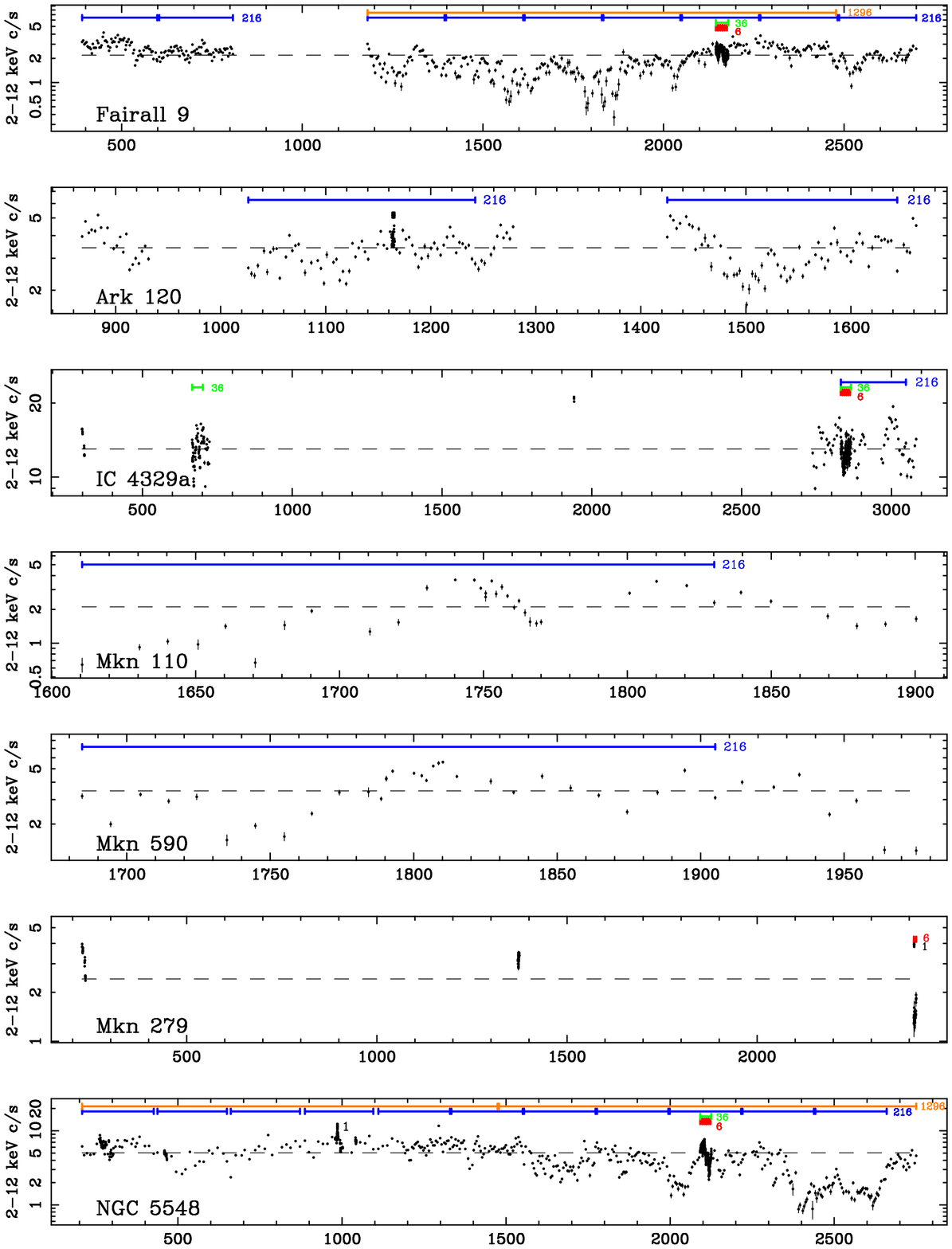}
\caption{Figure 1, cont'd.}
\end{figure}

\setcounter{figure}{0}
\figurenum{1}
\begin{figure}
\epsscale{0.94}
\plotone{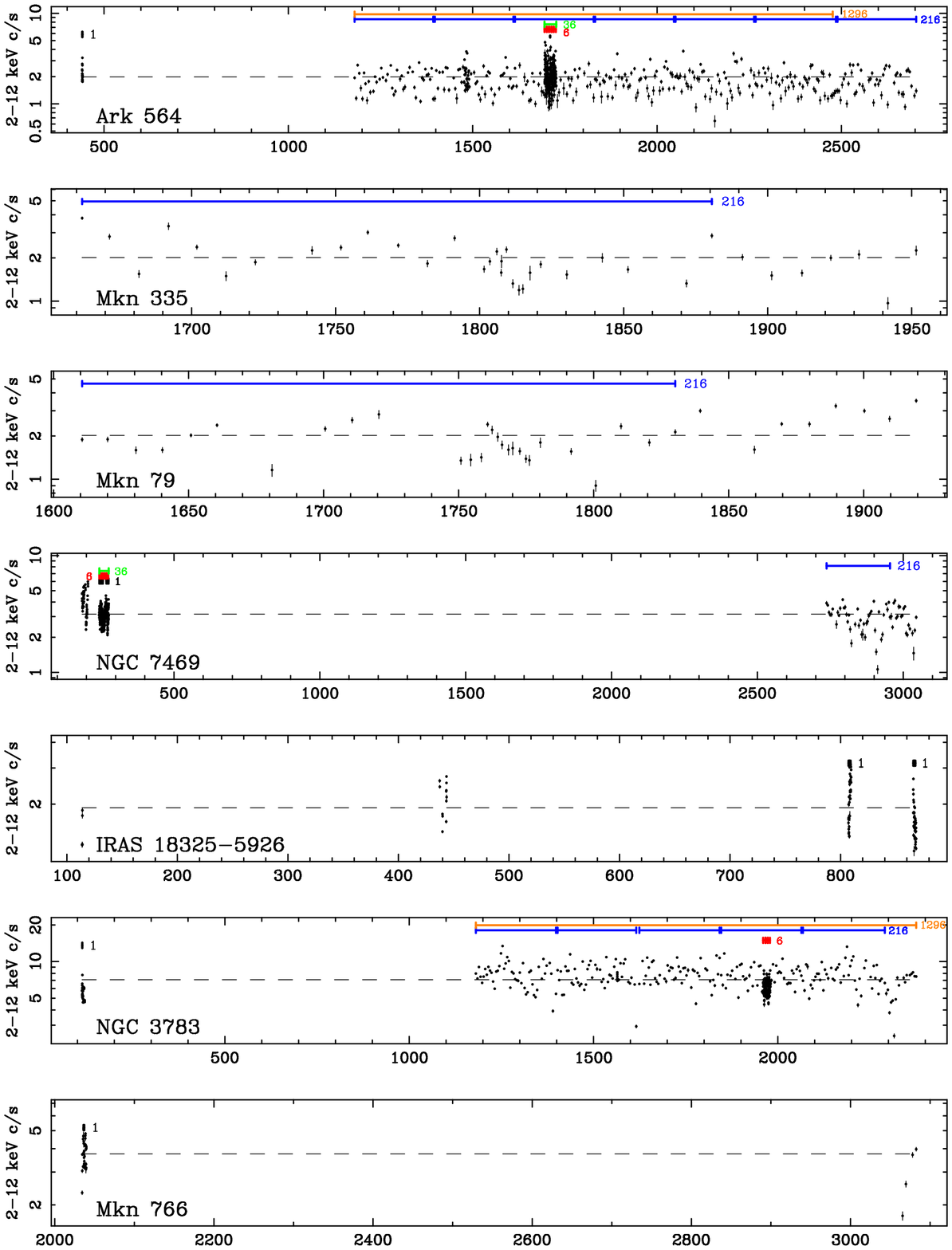}
\caption{Figure 1, cont'd.}
\end{figure}

\setcounter{figure}{0}
\figurenum{1}
\begin{figure}
\epsscale{0.94}
\plotone{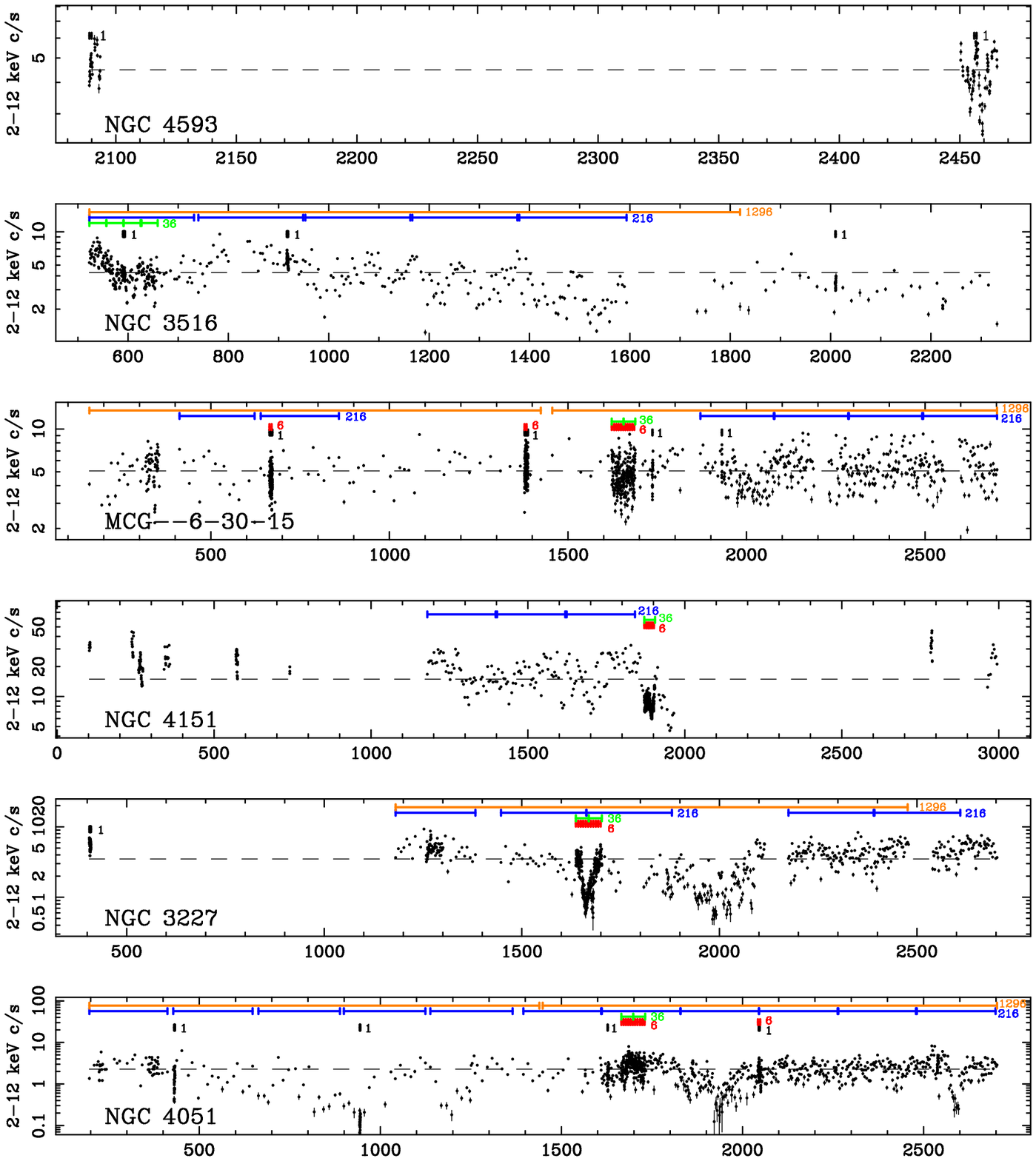}
\caption{Figure 1, cont'd.}
\end{figure}

\begin{figure}
\figurenum{2}
\epsscale{0.84}
\plotone{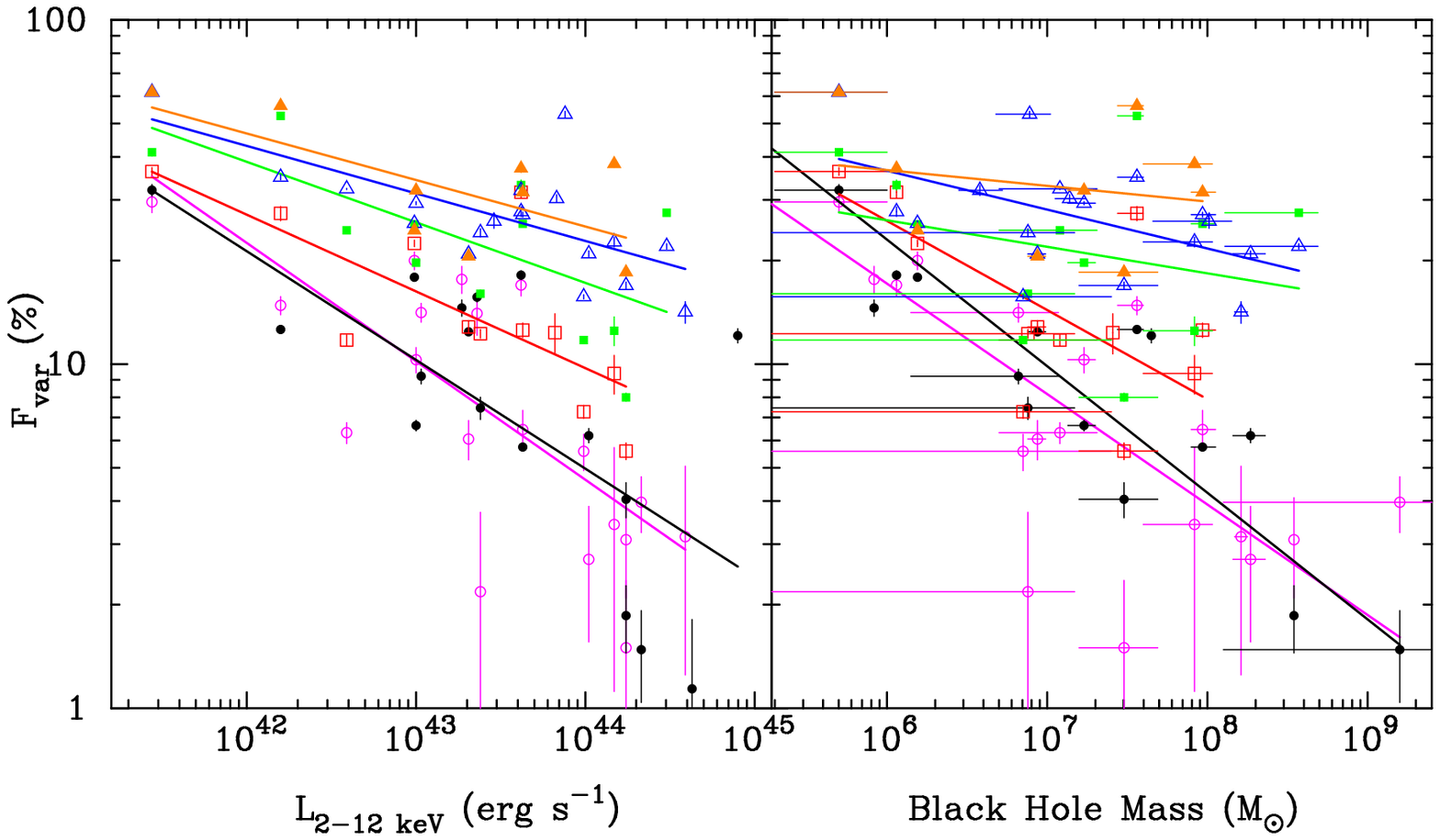}
\caption{$\fv$ plotted against 2--10 keV luminosity
(left) and black hole mass estimate $\mbh$ (right) for all time scales.
1 d {\it ASCA}, 1 d {\it RXTE}, 6 d 36 d, 216 d, and 1296 d
time scale data points are denoted by 
purple open circles,
black filled circles,
red open squares,
green filled squares,
blue open triangles,
and orange filled triangles, respectively.
The best-fit lines for each time scale are 
the same color.}
\end{figure}

\begin{figure}
\figurenum{3}
\epsscale{0.54}
\plotone{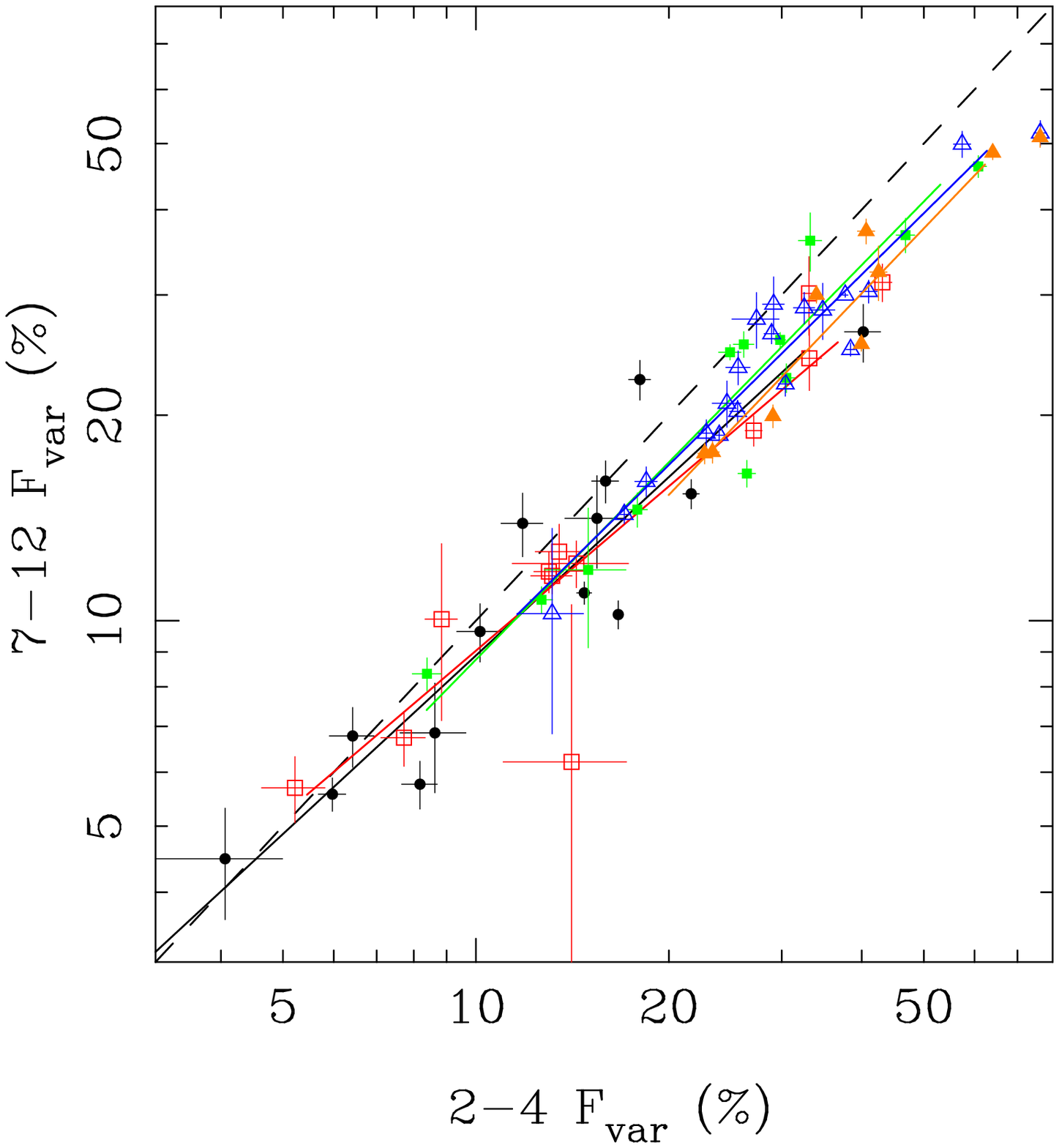}
\caption{7--12 keV $\fv$ plotted against 2--4 keV $\fv$.
Data points and best-fit lines are denoted the same as in Figure~2.
A source with equally strong variability in the two bands would lie on the
dashed line, but the vast majority of the light curves exhibit stronger
variability in the softer band.}
\end{figure}

\begin{figure}
\figurenum{4}
\epsscale{0.54}
\plotone{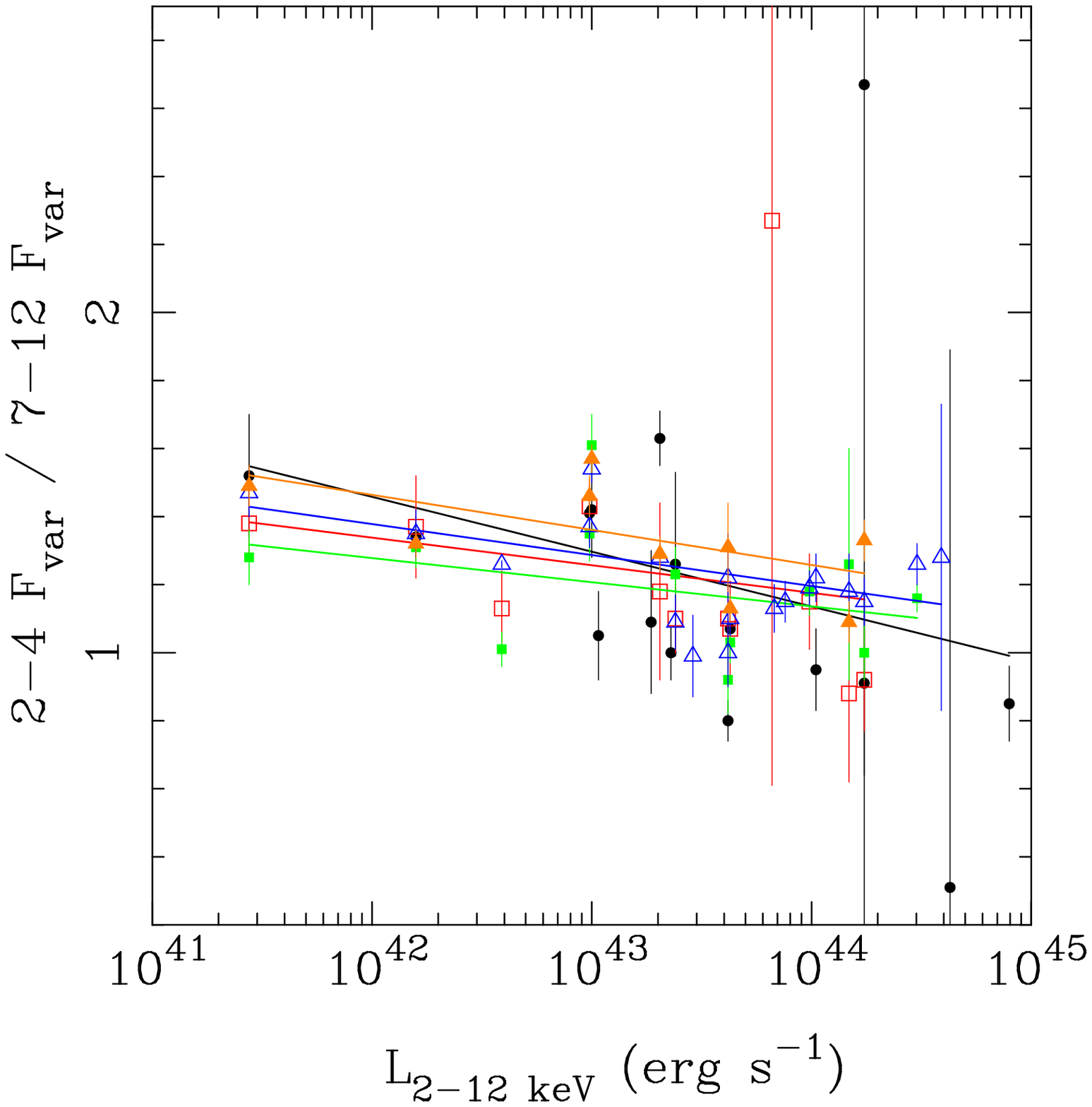}
\caption{The ratio of 2--4 keV $\fv$ / 7--12 keV $\fv$
plotted against source luminosity. Data points and 
best-fit lines are denoted the same as in Figure~2.
There is tentative evidence for relatively less luminous 
sources to display increasingly stronger variability in 
the soft band compared to the hard band.}
\end{figure}

\begin{figure}
\figurenum{5}
\epsscale{0.84}
\plotone{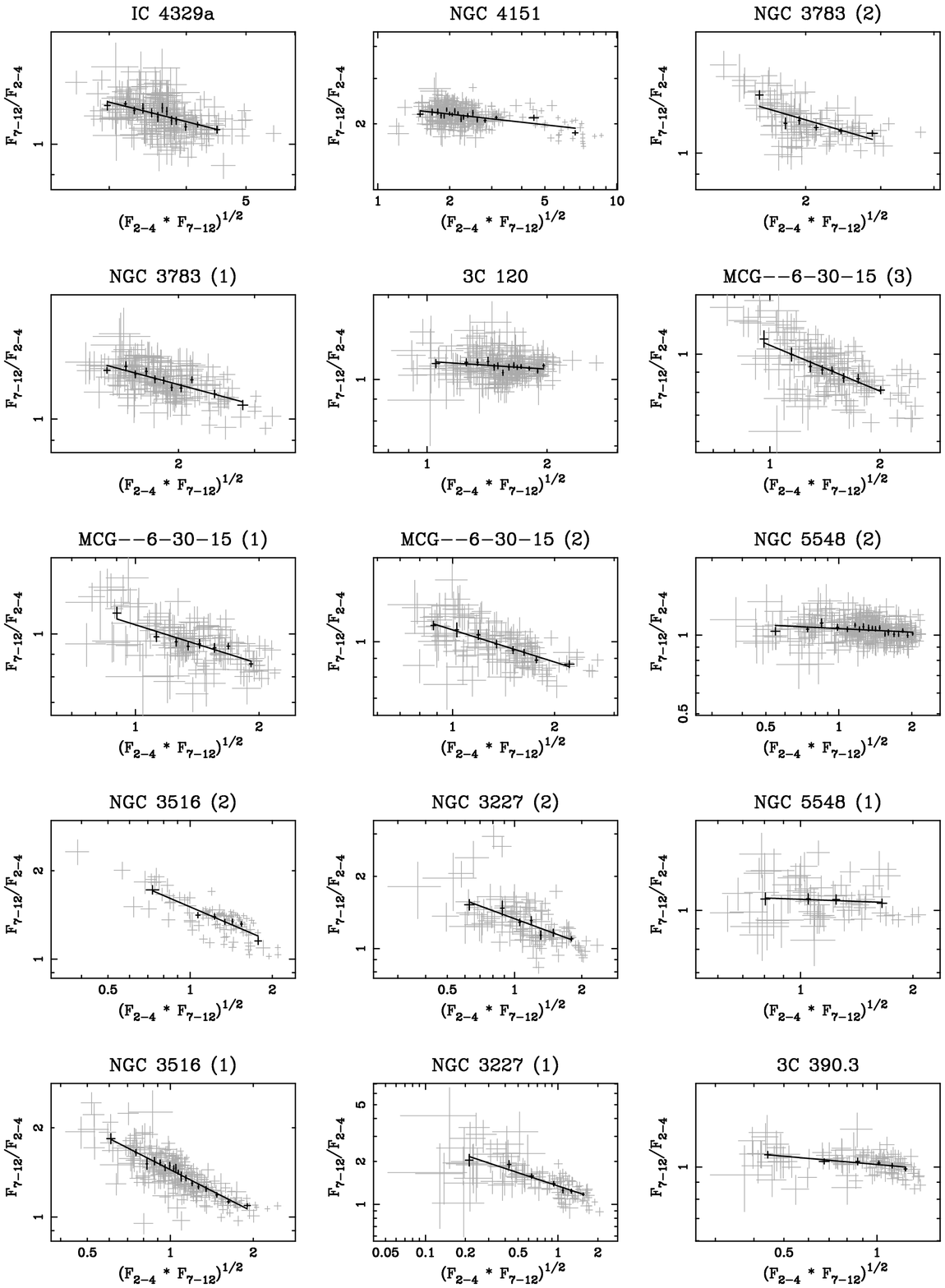}
\caption{Plots of the 7--12 keV/2--4 keV hardness ratio (HR)
against geometric mean count rate for the 27 light curves with 
adequate data. Gray error bars represent the unbinned data; 
black points represent the binned data. The solid lines are 
the best linear fits to the binned data.}
\end{figure}

\setcounter{figure}{4}
\begin{figure}
\figurenum{5}
\epsscale{0.84}
\plotone{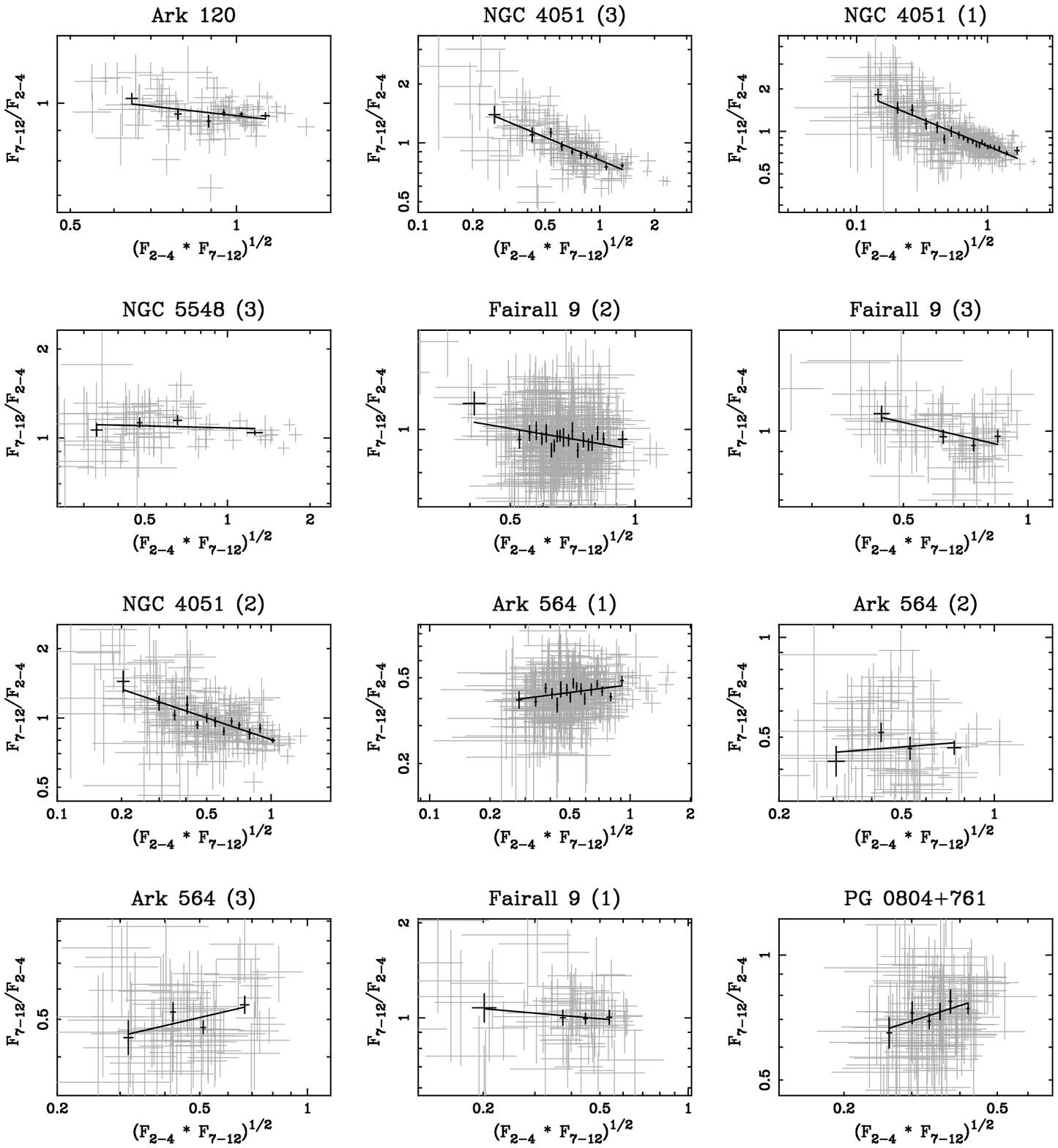}
\caption{Continued.}
\end{figure}

\begin{figure}
\figurenum{6}
\epsscale{0.94}
\plotone{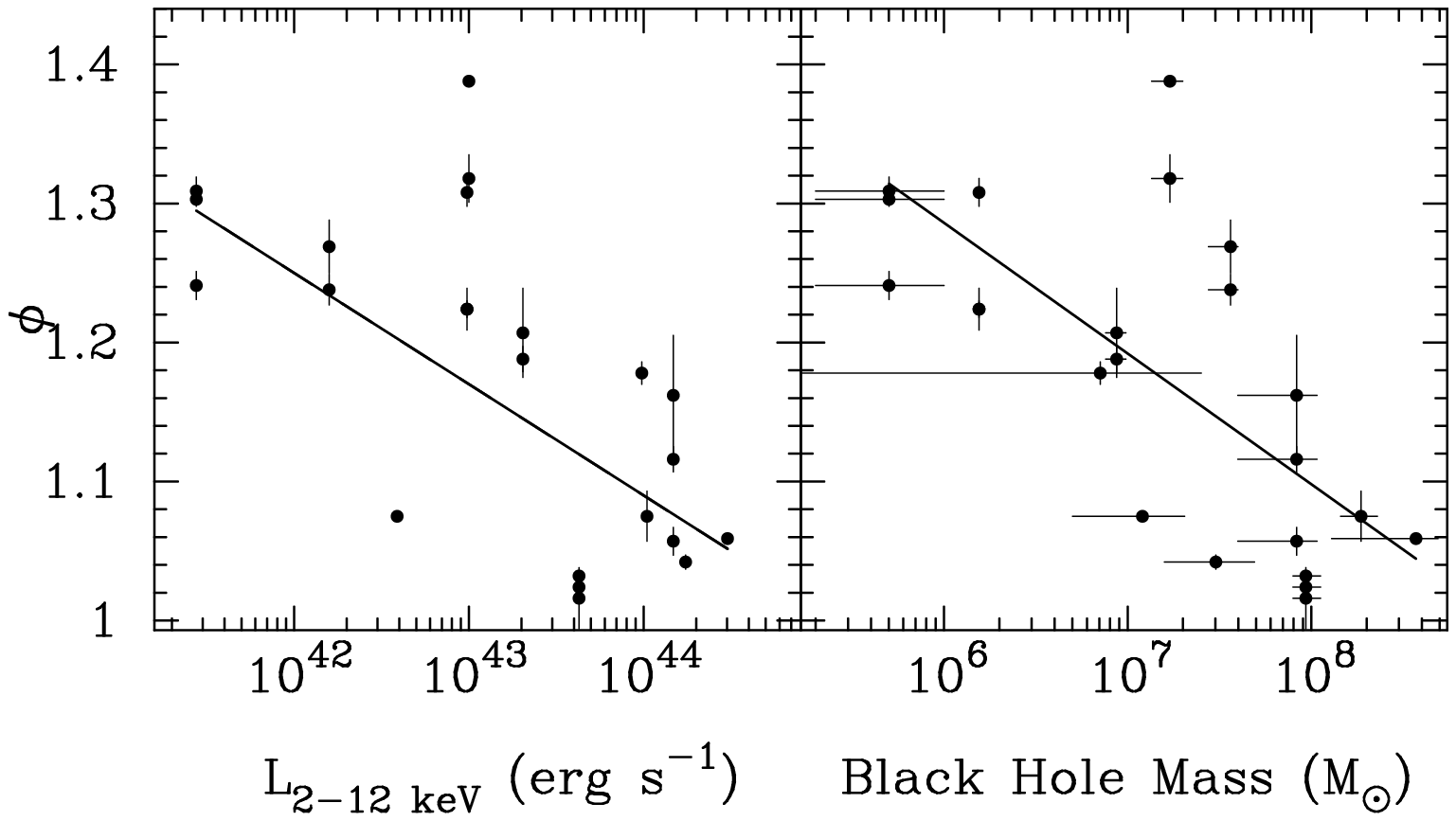}
\caption{The parameter $\phi$, which quantifies the decrease in 
7--12 keV/2--4 keV hardness ratio (HR) for every doubling in flux,
plotted against 2--12 keV luminosity and black hole mass estimate $\mbh$.
Ark~564 and PG~0804+761 are excluded.
There is a tendency for relatively less luminous or less massive
sources to display more overall spectral variability.}
\end{figure}

\begin{figure}
\figurenum{7}
\epsscale{0.83}
\plotone{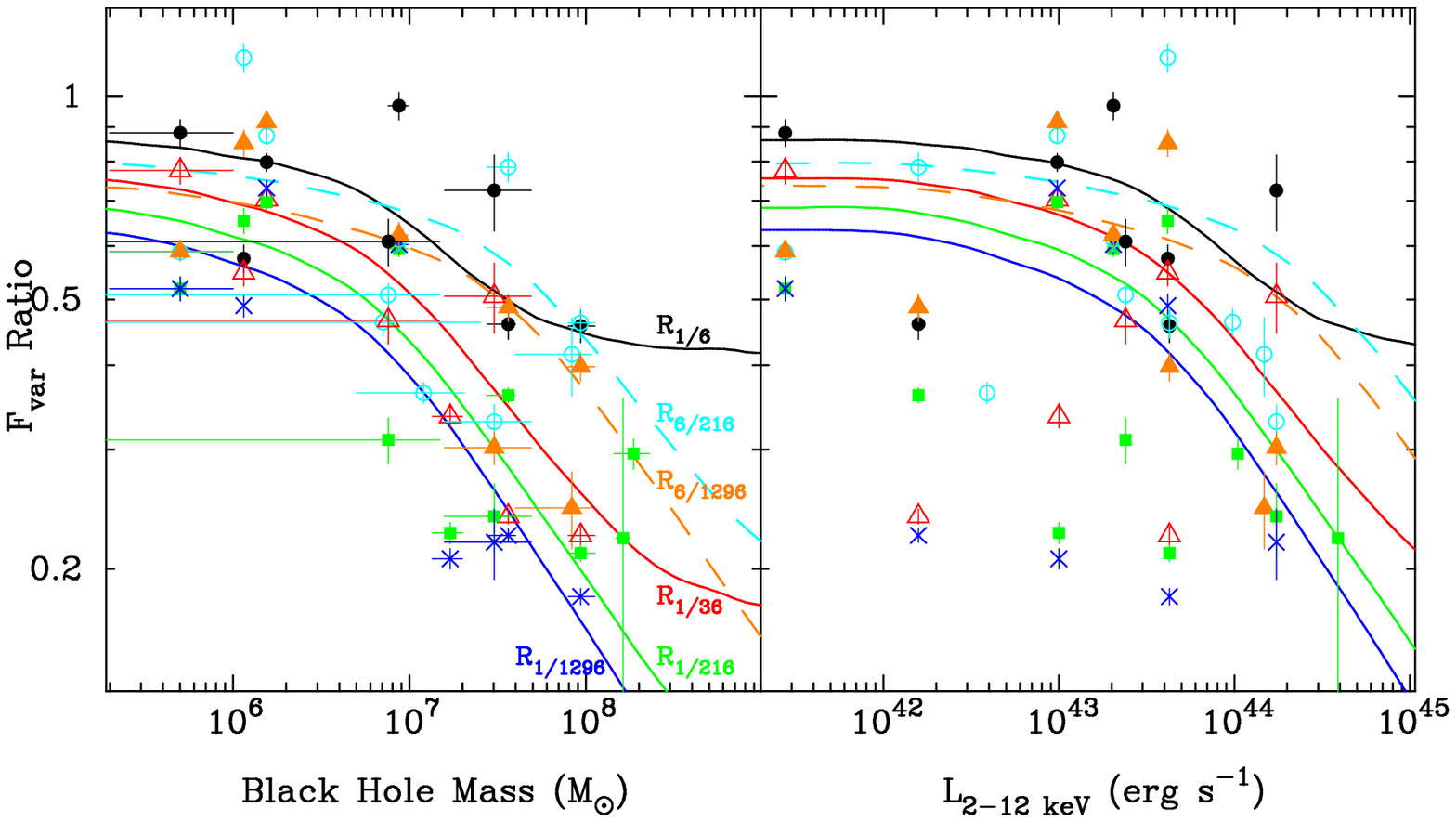}
\caption{Predicted and measured ratios of $\fv$ for six  
time scale combinations plotted against black hole mass estimate $\mbh$
(left) and 2--12 keV luminosity (right).
The solid lines and dotted lines denote the model $\fv$ ratios, 
derived from linearly scaling the PSD break
frequency with black hole mass using the best-fit relation
T$_b$=$\mbh$/10$^{6.7}$$\Msun$. The model ratios for 
the $\fv$ ratio --$L_{2-12}$ plot were derived from linearly 
scaling PSD break frequency with X-ray luminosity using the 
best-fit relation T$_b$=$L_{2-12~keV}$/10$^{43.5}$ erg~s$^{-1}$.
The black, red, green, and blue solid lines mark
the model $\fv$ ratios $R_{1/6}$, $R_{1/36}$, $R_{1/216}$ and $R_{1/1296}$,  
respectively, where e.g., $R_{1/6}$ denotes the ratio
$F_{var}$(1 d) / $F_{var}$(6 d).
The ratios $R_{6/216}$, $R_{6/1296}$
are marked by cyan and orange dotted lines, respectively.
Note that, as one scales a PSD towards higher temporal frequencies,
$\fv$ measured over a fixed time scale (fixed sampling
window in the frequency domain) will increase, but 
at a faster rate for relatively shorter sampling time scales.
The model ratios hence follow 
$R_{1/6}$  $>$ $R_{1/36}$ $>$ $R_{1/216}$ $>$ $R_{1/1296}$
(the black, red, green, and blue curves, respectively), 
$R_{6/36}$ $>$ $R_{6/216}$ $>$ $R_{6/1296}$
(not plotted, cyan, and orange, respectively), etc.
Note also that the $\fv$ ratio will tend to flatten off 
at high frequencies when both sampling time scales are 
much shorter than the PSD break time scale; i.e., this is why
$R_{1/6}$ flattens off at high mass/ low frequencies.
The measured ratios of $R_{1/6}$, $R_{1/36}$, $R_{1/216}$, $R_{1/1296}$, 
$R_{6/216}$, $R_{6/1296}$, are denoted by
black filled circles,  red open triangles, green filled squares, blue crosses,
cyan open circles and orange filled triangles, respectively.}
\end{figure}

\begin{figure}
\figurenum{8}
\epsscale{0.80}
\plotone{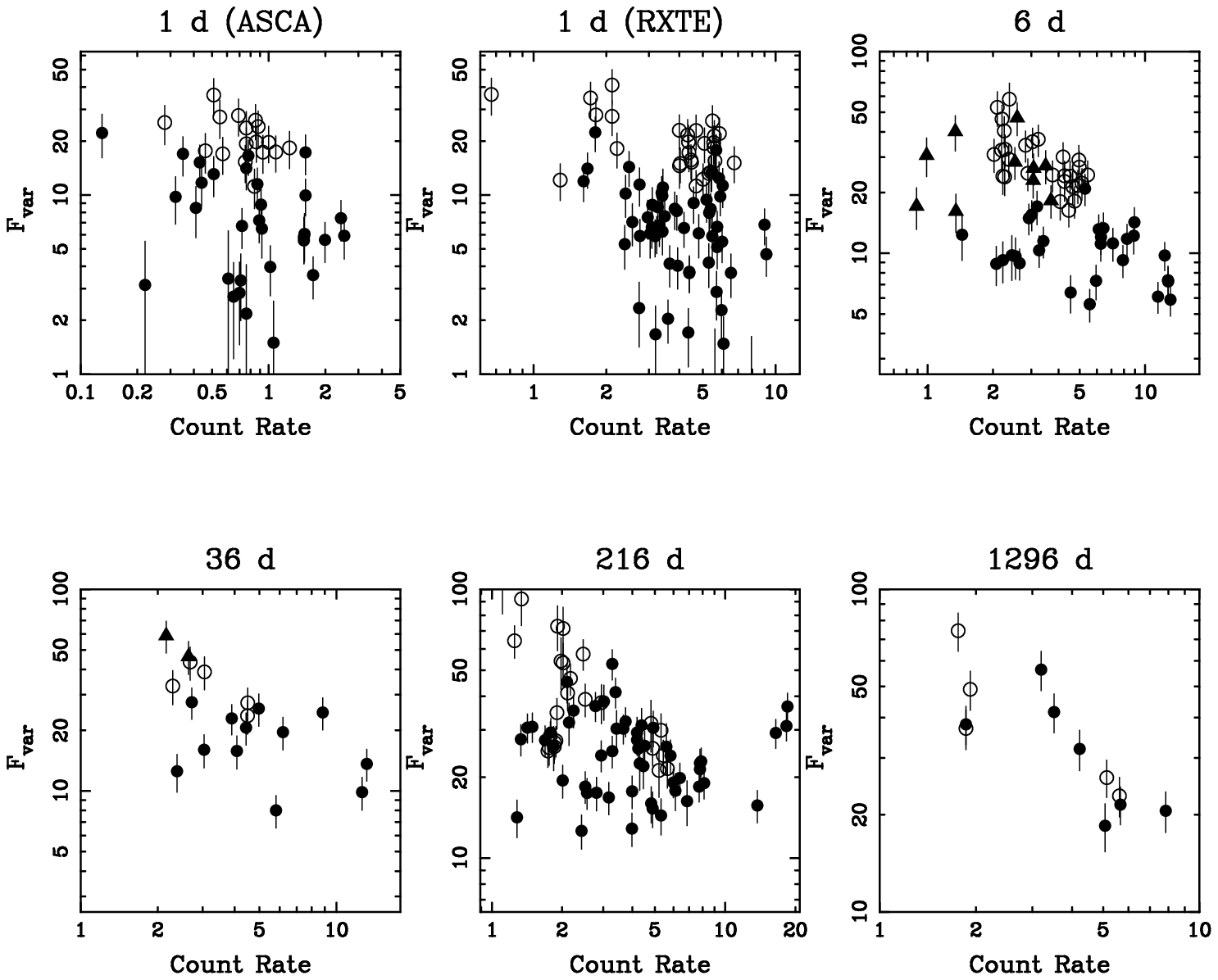}
\caption{Logarithm of 2--12 keV $\fv$ (2--10 keV for {\it ASCA}) 
plotted against the logarithm of the count rate.
Filled circles denote broad-line
Seyferts. Open circles denote narrow-line/ soft spectrum Seyferts,
which tend to be more variable and less luminous in the \xte\ bandpass.
Filled triangles in the 6 d and 36 d plots are NGC~3227; excluding
those data points, no trends with count rate are evident 
for the broad-line sources overall.}
\end{figure}
 
\begin{figure}
\figurenum{9}
\epsscale{0.80}
\plotone{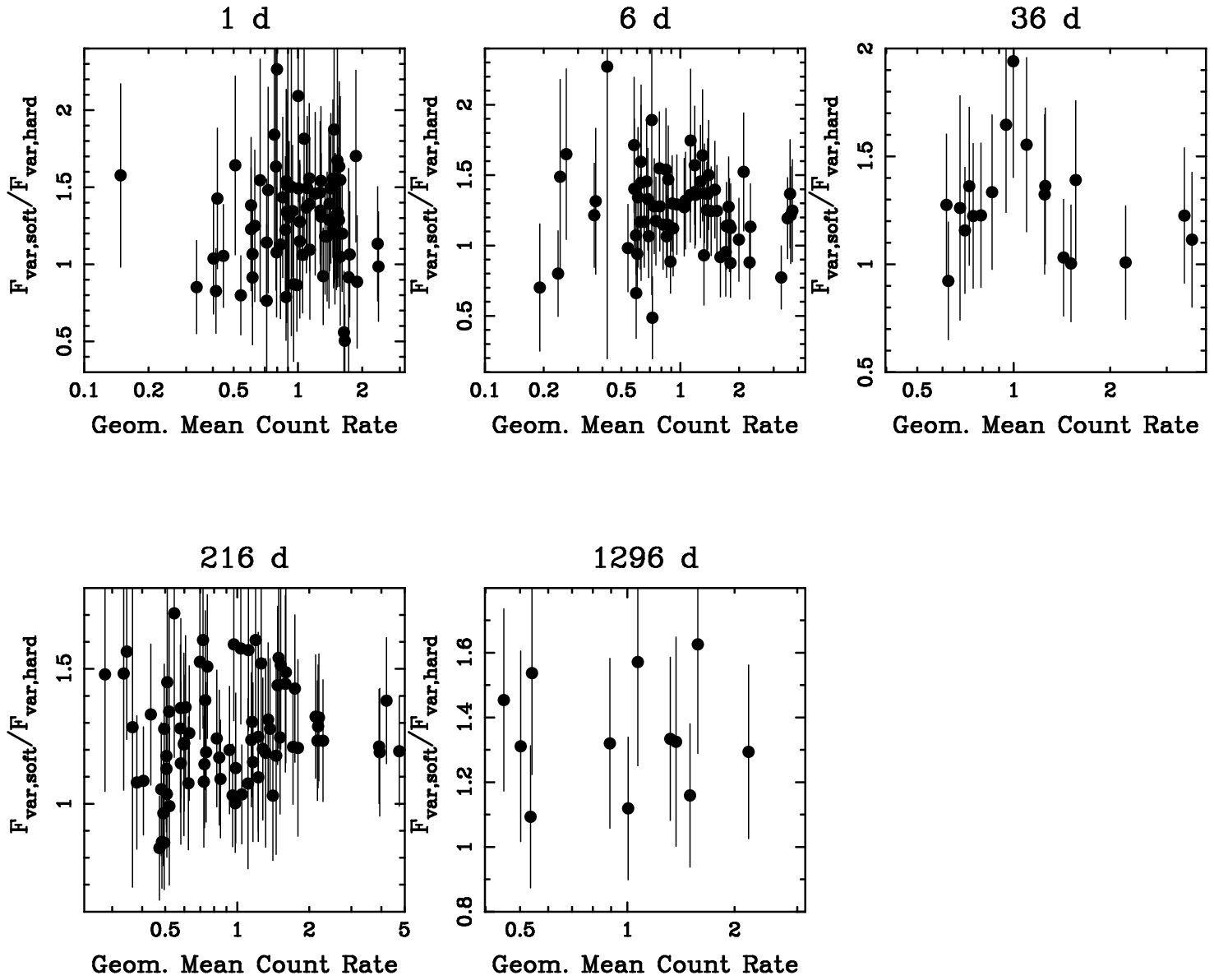}
\caption{$F_{var,soft}$/$F_{var,hard}$ plotted against the logarithm of the count rate. No trends with count rate are evident.}
\end{figure}
 
\begin{figure}
\figurenum{10}
\epsscale{0.40}
\plotone{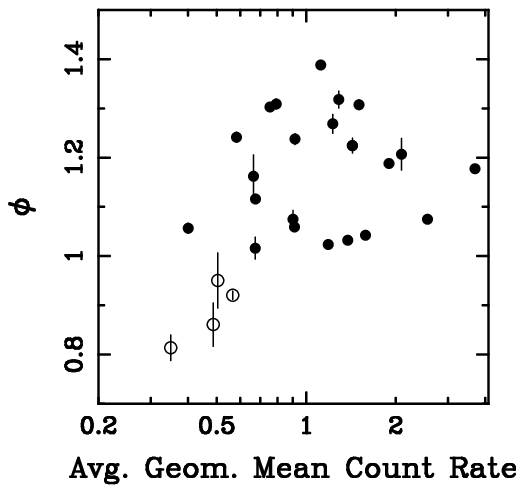}
\caption{Plot of the parameter $\phi$ against average geometric mean count rate.
Open circles denote Ark~564 and PG~0804+761;
their low values of $\phi$ may be an artifact of the low count rates and systematic
errors in the PCA background model.
For the other sources, $\phi$ is independent of count rate.}
\end{figure}

\end{document}